\documentclass[preprint,12pt,3p]{elsarticle}

\usepackage{amssymb}
\usepackage{booktabs}
\usepackage{appendix}
\usepackage{CJK}
\usepackage{stfloats}
\usepackage{amsmath,amssymb,amsfonts}
\usepackage{algorithmic}
\usepackage{graphicx}
\usepackage{textcomp}
\usepackage{xcolor}
\usepackage{algorithm} 
\usepackage{multirow} 
\usepackage{diagbox}
\usepackage{algorithmic}
\usepackage{caption}
\usepackage{makecell}

\journal{Nuclear Physics B}

\begin{document}

\begin{frontmatter}
	
	\title{Detecting Unknown HTTP-based Malicious Communication Behavior via Generated Adversarial Flows and Hierarchical Traffic Features}
	
	\address[label1]{National Computer Network Emergency Response Technical Team/Coordination Center of China, Beijing, China}
	\address[label2]{Institute of Information Engineering, Chinese Academy of Sciences, Beijing, China}
	\address[label3]{Key Laboratory of Network Assessment Technology, University of Chinese Academy of Sciences, Beijing, China}
	\address[label4]{School of Cyber Security, University of Chinese Academy of Sciences, Beijing, China}

	\author[label1]{Xiaochun Yun}
	\ead{yunxiaochun@cert.org.cn}
	
	\author[label2,label4]{Jiang Xie}
	\ead{xiejiang@iie.ac.cn}
	
	\author[label2,label3,label4]{Shuhao Li\corref{cor1}}
	\cortext[cor1]{The corresponding author of this paper is Shuhao Li.}
	\ead{lishuhao@iie.ac.cn}
	
	\author[label2,label3,label4]{Yongzheng Zhang}
	\ead{zhangyongzheng@iie.ac.cn}
	
	\author[label2,label4]{Peishuai Sun}
	\ead{sunpeishuai@iie.ac.cn}
	
	\begin{abstract}
		Malicious communication behavior is the network communication behavior generated by malware (botnet, spyware, \textit{etc.}) after victim devices are infected. Experienced adversaries often hide malicious information in HTTP traffic to evade detection. 
		However, related detection methods have inadequate generalization ability because they are usually based on artificial feature engineering and outmoded datasets. 
		In this paper, we propose an HTTP-based Malicious Communication traffic Detection Model (HMCD-Model) based on generated adversarial flows and hierarchical traffic features. HMCD-Model consists of two parts.
		The first is a generation algorithm based on WGAN-GP to generate HTTP-based malicious communication traffic for data enhancement.
		The second is a hybrid neural network based on CNN and LSTM to extract hierarchical spatial-temporal features of HTTP-based traffic. 
		In addition, we collect and publish a dataset, HMCT-2020, which consists of large-scale malicious and benign traffic during three years (2018-2020). 
        Taking the data in HMCT-2020(18) as the training set and the data in other datasets as the test set, the experimental results show that the HMCD-Model can effectively detect unknown HTTP-based malicious communication traffic.
		It can reach \textit{F}1 $\approx $ 98.66\% in the dataset HMCT-2020(19-20), \textit{F}1 $\approx $ 90.69\% in the public dataset CIC-IDS-2017 and \textit{F}1 $\approx$ 83.66\% in the real traffic,  which is 20+\% higher than other representative methods on average.
		This validates that HMCD-Model has the ability to discover unknown HTTP-based malicious communication behavior.
	\end{abstract}
	
	\begin{keyword}
		Malicious Behavior Detection \sep CNN \sep LSTM \sep GAN \sep Hierarchical Features
	\end{keyword}
	
\end{frontmatter}


\section{Introduction}




Malicious traffic detection generated by malware is one of the hot issues in cyber security \cite{ghafir2015survey, jose2018survey}. 
In this paper, we call the network communication behavior generated by malware (botnet, spyware, \textit{etc.}) after malware infects the victim device as \textit{malicious communication behavior}. 
One of the main carriers of these behavior is HTTP traffic\cite{wang2017detecting}. 
Experienced adversaries construct HTTP-based malicious communication traffic by imitating the network behavior of benign users and hiding malicious information into the fields used commonly in benign traffic. 
These unknown HTTP-based malicious traffic is highly similar to benign traffic,  usually can bypass the detection systems.
Therefore, it is difficult but necessary to detect HTTP-based malicious communication behavior, especially unknown. This motivates researchers to pursue advanced detection techniques.

The key to detecting unknown HTTP-based malicious communication behavior is to improve the generalization ability of detection methods, i.e., the ability to discover unknown attacks by the known. There are two main challenges for detecting: 

1) \textit{\textbf{Feature extract}}. Traffic feature of HTTP-based malicious communication behavior is complex. However, many detection methods (\cite{wang2017detecting, salo2019dimensionality}, \textit{etc.}) rely on feature rules and expert knowledge so that they are difficult to grasp the essential laws of HTTP-based malicious communication behavior.
The establishment of artificial feature engineering in a single experimental environment will cause the model to over-fit, which limits the generalization of method. This makes it difficult for a method to detect unknown HTTP-based malicious communication traffic. 
Namely, a method may perform well in assigned datasets, but poor in other large-scale datasets and real traffic environments. 

2) \textit{\textbf{Experimental dataset}}. The data scale associated with HTTP-based malicious communication behavior is relatively small. Many works (\cite{aburomman2017survey, selvakumar2019firefly}, \textit{etc.}) rely on private experimental data (the collection period is short and the collection points are limited), or rely on public malicious traffic datasets. 
However, due to the issue of timeliness, it is difficult for those datasets to cover all forms of such malicious traffic that occurs in the future. 

In this paper, to cope with the above challenges in the detection of unknown HTTP-based malicious communication behavior, we propose an HTTP-based Malicious Communication traffic Detection Model (HMCD-Model). The main contributions are as follows:

\begin{itemize}
	\item \textbf{We analyze the HTTP-based malicious communication behavior from the perspective of adversary.} We employ WGAN-GP \cite{gulrajani2017improved} to synthesize Generated Adversarial Flows (GAFs) with maliciousness, compliance, covertness and multiformity for data enhancement. GAFs look highly similar to benign flows, which can be used as a supplement to labeled data and improve the generalization of HMCD-Model.
	
	\item \textbf{We propose a prototype system composed of a hybrid neural network to extract the hierarchical spatial-temporal features of HTTP traffic from packet level and flow level.} In this system, CNN is used to extract the spatial features of a packet, and LSTM is used to extract the temporal features of a flow. 
	In addition, statistical features are also extracted hierarchically to improve detection performance. 
	And we discard miss-leading attributes (ip, url, \textit{etc.}) that could easily lead to misjudgment in data pre-processing.
	
	\item \textbf{We publish a dataset HMCT-2020 based on the real network environment}, which consists of large-scale HTTP-based malicious communication traffic and benign traffic during three years (2018-2020)\footnote{The published dataset can be found at \textit{https://github.com/BitBrave-Xie/HMCD-Model}.}.
	There are about 76,760 malicious flows and 4,798,110 benign flows.
	HMCT-2020 can not only support our experiments but also help researchers further study  HTTP-based malicious communication behavior. 
\end{itemize}

Experimental results show that HMCD-Model has excellent detection performance. 
In HMCT-2020 dataset, $ F1 $ is 99.46\%(+0.19, -0.30), and $ FPR $ is 0.48\%(+0.20, -0.34). For generalization (Taking the data in HMCT-2020(18) as the training set and the data in other datasets as the test set), our model has obvious advantages compared with the representative works in HMCT-2020 and the public dataset CIC-IDS-2017\cite{sharafaldin2018toward}.
In addition, we also collect malicious traffic generated by malware and a large amount of benign background traffic from the real world,
In the comparative experiment, the $ F1 $ and $ FPR $ of HMCD-Model can reach 83.66\%(+2.15, -3.79) and 2.57\%(+3.26, -1.71), which are also better than baselines and other methods\cite{wang2017detecting, salo2019dimensionality}. The results prove that our method has a stronger ability to discover unknown malicious communication behavior.

The remainder of this paper is organized as follows. Section 2 introduces the related work. In Section 3, we analyze the HTTP-based malicious communication behavior. Feature analysis of HTTP traffic is in Section 4. Subsequently, Section 5 introduces the composition of HMCD-Model. In Section 6, we evaluate our method and show the relevant experimental results. Finally, we discuss and summarize in Section 7 and Section 8, respectively.

\section{Related Work}
In cyber security, malicious behavior detection is a hot issue\cite{jose2018survey}. 
We detect unknown HTTP-based malicious communication behavior based on deep learning, which belongs to the field of intrusion detection.
Next, the related research status will be introduced.

\subsection{Malicious Behavior Detection}
Malicious behavior detection methods used in Intrusion Detection Systems (IDSs) \cite{maki1989intrusion} can be divided into feature detection and anomaly detection.
Feature detection \cite{cannady1998artificial}, also called misuse detection, fits the behavior patterns of known attacks and establishes a corresponding feature behavior database.
Anomaly detection \cite{chandola2009anomaly}, also called behavior detection, mainly builds a feature database by fitting the characteristics of benign network behavior.
Anomaly detection is slightly weaker than feature detection when detecting known attacks. However, anomaly detection can detect 0-day attacks more effectively. And this is very important for network security, because the network environment is becoming more and more complex, new 0-day attacks are constantly occurring, and a method that can effectively detect new attacks is necessary.
%

\subsubsection{Malicious Behavior Detection Based on Feature Selection and Machine Learning}
Many works are based on some public network datasets (ISCX-2012 \cite{shiravi2012toward}, KDD CUP 99 \cite{stolfo1999kdd}, CIC-IDS-2017, \textit{etc.}), and then malicious behavior detection methods combining feature selection and machine learning are employed.

Aburomman \emph{et al.} \cite{aburomman2017survey} review intrusion classification algorithms based on commonly used methods in the field of machine learning. In particular, considering integration methods of homogeneous and heterogeneous types, various integration and hybrid technologies are studied.

Wang \emph{et al.} \cite{wang2017detecting} propose an effective and automatic malware detection method using the text semantics of network traffic, treating each HTTP flow generated by software as a text document and processing it through N-gram to extract text-level features. Then, an automatic feature selection algorithm based on chi-square test is used to identify meaningful features, and these features are used to establish a support vector machine (SVM) classifier \cite{suykens1999least} for malicious behavior detection.

Salo \emph{et al.} \cite{salo2019dimensionality} propose an ensemble classifier based on SVM, Instance-based learning algorithms (IBK) \cite{altman1992introduction}, and multilayer perceptron (MLP) \cite{white1963principles} for intrusion detection, which combines the approaches of Information Gain (IG) and Principal Component Analysis (PCA). The experimental results in datasets ISCX 2012, NSL-KDD \cite{tavallaee2009detailed} and Kyoto 2006+ \cite{song2011statistical} show that IG-PCA-Ensemble can learn more key features, and its classification accuracy, detection rate and false alarm rate are better than most of the existing advanced methods.

There are also many other studies about malicious behavior detection.
Some methods are mainly based on collected experimental datasets.
Wang \emph{et al.}\cite{wang2020botmark} propose BotMark for botnets detection based on flow-based and graph-based network traffic behavior.
Du \textit{et al.} \cite{du2018network} use SVM to differentiate the two types of anomaly in the mixed traffic.
Shrestha \emph{et al.} \cite{shrestha2015support} use SVM for covert channel detection.
Others methods are mainly based on public intrusion detection datasets.
Zhou \emph{et al.} \cite{zhou2020building} propose a heuristic dimension reduction algorithm CFS-BA-Ensemble based on feature selection and ensemble learning technology.
Selvakumar \emph{et al.} \cite{selvakumar2019firefly} deploy filters and wrappers based on the firefly algorithm in the feature selector, and use C4.5 and BN to classify in KDD CUP 99 dataset.
Hajisalem \emph{et al.} \cite{hajisalem2018hybrid} propose a hybrid classification method based on Artificial Bee Colony (ABC) \cite{karaboga2008performance} and Artificial Fish Swarm (AFS) \cite{gupta2019efficient} algorithms. The method is superior to traditional machine learning methods in the NSL-KDD dataset and the UNSW-NB15 dataset \cite{moustafa2015unsw}.

\subsubsection{Malicious Behavior Detection Based on Deep Learning}
Deep learning is widely researched and applied in the field of intrusion detection due to its powerful feature extraction capabilities \cite{kwon2019survey}. After the traffic is simply pre-processed, the neural network can automatically extract features and do not require researchers to put more effort into establishing feature engineering.

Chowdhury \textit{et al.} \cite{chowdhury2017few} extract outputs from different layers in CNN and implement a linear SVM and 1-nearest neighbor classifier for few-shot intrusion detection.
Kim \emph{et al.} \cite{kim2018multimodal} propose a multi-mode deep learning method for malware detection.
Caviglione \emph{et al.} \cite{caviglione2015seeing} use neural networks and decision trees to detect malware using covert channels. Du \emph{et al.} \cite{du2017deeplog} use LSTM to perform malicious behavior detection.

\subsection{Traffic Generation based on GAN}
Currently, many researchers use deep learning technology to synthesize traffic data, which is used to bypass detection systems or enrich experimental data. Attempts to introduce the GAN to this field have shown promise. Zingo \emph{et al.} \cite{zingo2020can} propose the "GAN vs Real (GvR) score", a task-based metric which quantifies how well a traffic GAN generator informs a classifier compared to the original data. Experiments show that it is possible to train accurate traffic anomaly detectors with GAN-generated network traffic data based on GvR.

Li \emph{et al.} \cite{li2019dynamic} propose a dynamic traffic camouflaging technique, coined FlowGAN, to dynamically morph traffic feature as another "normal" network flow to bypass Internet censorship. The core idea of FlowGAN is to automatically learn the features of the "normal" network flow, and dynamically morph the on-going traffic flows based on the learned features by GAN. Experimental results on a dataset involving 10,000 realworld flows show that the effectiveness and the efficiency of FlowGAN.

Ring \emph{et al.} \cite{ring2019flow} generate flow data based on GANs and propose three different flow-based data pre-processing methods in order to convert them to continuous values. On this basis, a network traffic evaluation method based on domain knowledge definition quality test is proposed. Experiments on the CIDDS-001 dataset \cite{ring2017flow} show that two of the three methods can generate high-quality data.


Lin \emph{et al.} \cite{lin2018idsgan} propose a framework based on GANs, namely IDSGAN, to generate adversarial samples to evade the detection of IDSs. IDSGAN uses generators to convert original malicious traffic into hostile malicious traffic and uses discriminators to simulate a black box detection system.

Cheng \emph{et al.} \cite{cheng2019pac} propose a GAN method for creating network traffic data at the ip packet layer. It prove feasibility in the generation of real traffic flows such as ICMP Pings, DNS queries, and HTTP web requests. Experiments show that the generated packets can be successfully transmitted through the Internet and the corresponding response can be obtained.


Jan \emph{et al.} \cite{jan2020throwing} propose request data synthesis method to synthesize unseen (or future) robot behavior distribution. The synthesis method has distributed perception capabilities and uses two different generators in GAN to generate data for clustering regions and outliers in the feature space. 

Hao \emph{et al.} \cite{hao2021producing} propose a GAN-based data augmentation method. The features of flow-based network traffic are first preprocessed to fit the GAN, and then the Earth-Mover (EM) distance is employed to capture the distribution of low-dimensional subspace data, while an encoder structure is added to learn latent space representations to enhance the vanilla GAN. They construct an imbalanced dataset based on a real-world dataset and compare it with other methods, obtaining better performance in terms of recall, F1 score and AUC.

Cheng \emph{et al.} \cite{cheng2021packet} propose Attack-GAN based on the structure of SeqGAN \cite{yu2017seqgan}, to generate domain-constrained adversarial network traffic at the packet level.
Specifically, adversarial packet generation is formulated as a sequential decision process.
In this case, each byte in the packet is considered a token in the sequence.
The generator's goal is to choose a token that maximizes its expected final reward.
Generated network traffic and benign traffic are classified by black box IDS.
The prediction results of IDS are fed into the discriminator to guide the update of the generator.
Experimental results verify that the generated adversarial examples are able to deceive many existing black-box IDSs.

\subsection{Analysis and Summary}
For malicious behavior detection, there are two main challenges in the above methods for detecting unknown HTTP-based malicious communication behavior based on the discussions above methods. First, these methods have a valuable reference for feature modeling, but it is difficult to directly apply to the feature extracting of HTTP-based malicious communication behavior under adversarial conditions. For instance, some unstable features (ip, url, \textit{etc.}) used before may be invalid. Moreover, hierarchical spatial-temporal features have not been considered too much. Second, these methods are usually tested on a single small-scale dataset, which cannot fully verify their generalization ability.

For traffic generation based on GAN, current methods for generating traffic samples are basically imitating specific datasets (CIDDS-001, \textit{etc.}) or specific formats to generate various flow-based statistical data (number of packets, duration of the flow, \textit{etc.}), not to generate real traffic that can be transmitted on the network, especially HTTP-based malicious communication behavior. This makes various traffic generation technologies have great limitations and can only be applied to a specific experimental scene.

Therefore, we build a hybrid neural network model for HTTP-based malicious communication behavior detection. Then, we propose a generation algorithm with good generality to synthesize GAFs. 
In addition, we construct a well-represented HTTP-based malicious communication traffic dataset to verify our model under different experimental conditions. 

\section{Analysis on HTTP-based Malicious Communication Behavior}

\begin{figure}[htbp]
		\vspace{-1.5em}
		\centering
		\centerline{\includegraphics[scale=0.53]{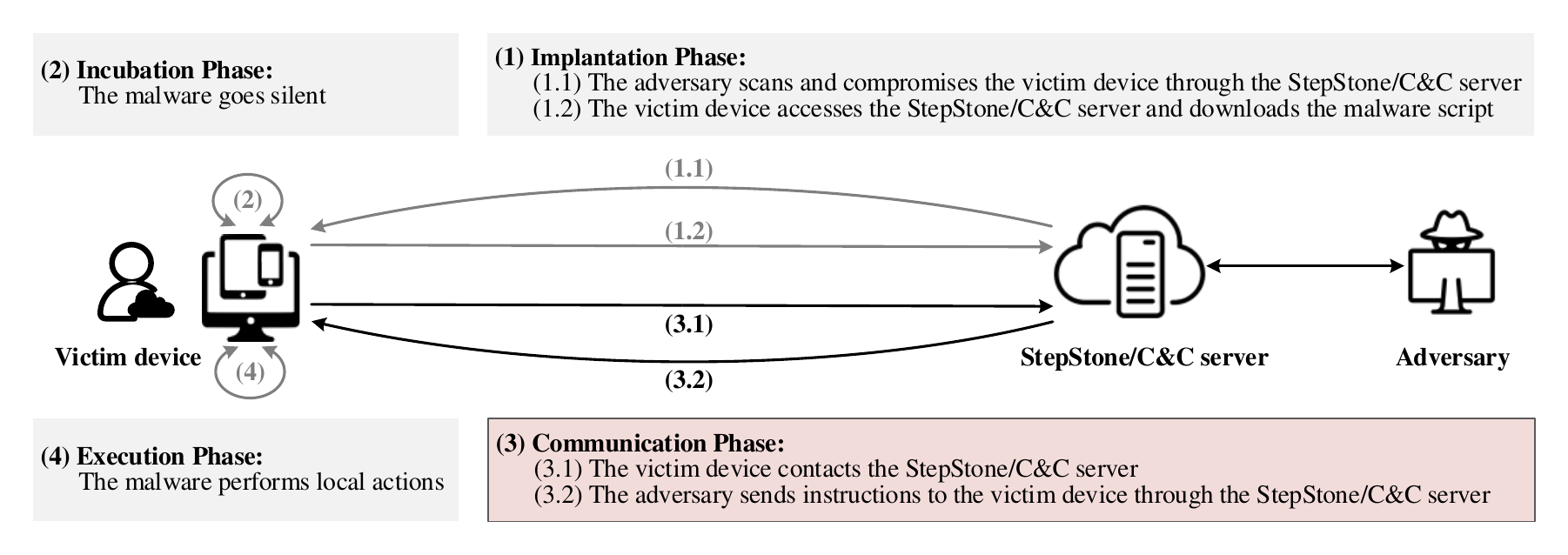}}
		\vspace{-1em}
		\caption{General process of HTTP-based malicious communication behavior.}
		\label{covert-attack}
		\vspace{-1em}
	\end{figure}

	We show the general process of HTTP-based malicious communication behavior, as shown in Fig.~\ref{covert-attack}. An HTTP-based malware attack can be divided into four phases: implantation phase, incubation phase, communication phase and execution phase. 
	1) During the implantation phase, the adversary scan the victim device, or the victim accesses the StepStone/C\&C server. 
	Then, an HTTP-based malware script is covertly downloaded to the victim device. 
	2) During the incubation phase, usually, the malicious script sleeps for a period of time and enters the incubation phase to avoid being detected. 
	3) Then, the malware script enters the communication phase and sends on-line packets to contact the StepStone/C\&C server. 
	4) Finally, the malware script enters the execution phase and begin to carry out various local actions according to instructions of the adversary.
	

	HTTP-based malicious communication behavior can be effectively detected in the communication phase according to the instruction and the type of malicious interaction, because this is the phase where the characteristics of malicious communication behavior are most obvious.
	During the communication phase, defenders can analyze the traffic generated by malware, then, detect HTTP-based malicious communication behavior and find the adversary. 
	
\begin{figure}[ht]
		\centerline{\includegraphics[scale=0.45]{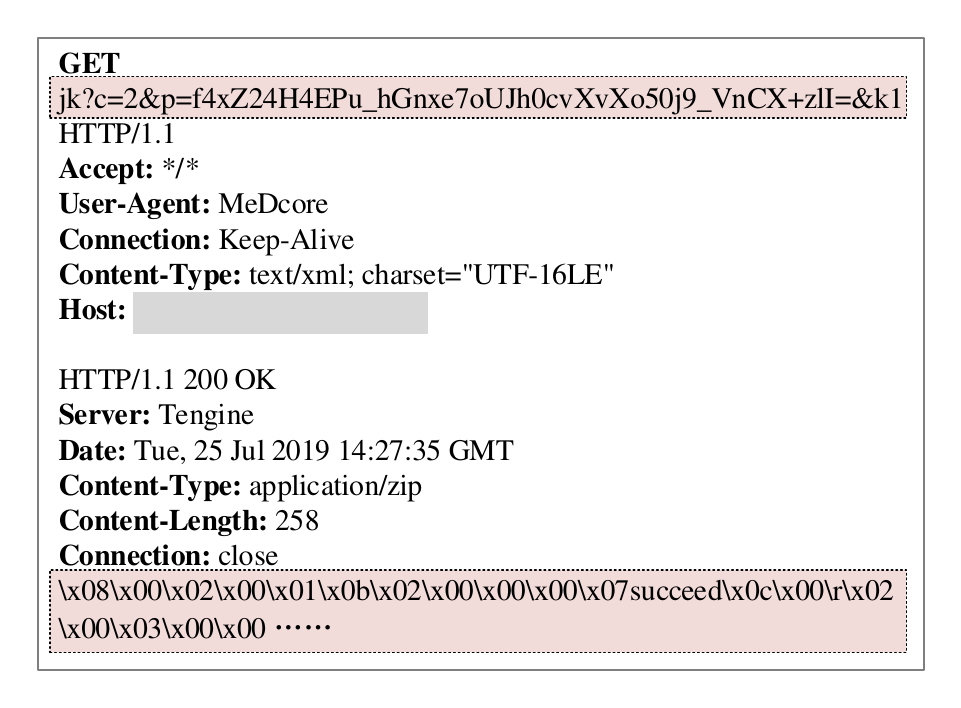}}
		\caption{A malicious flow consisting of two packets (a request and a response) generated by HTTP-based malicious communication behavior during the communication phase.}
		\label{attack-demo}
	\end{figure}

	However, experienced adversaries usually construct HTTP-based malicious communication traffic by imitating the network behavior of benign users to evade the detection of defenders, making it difficult to distinguish from benign traffic.
	For instance, fig.~\ref{attack-demo} shows a flow consisting of two packets that belongs to HTTP-based malicious communication behavior.
	And we can divide the packet into three different components: 
	1) \textit{Malicious components}, contain malicious content and has actual attack significance. For instance, the part enclosed by dotted line “jk?c=2\&=…” in Fig.~\ref{attack-demo}; 
	2) \textit{Fixed components}, are the components with a fixed format that may not exist in an HTTP packet but cannot be changed; 
	3) \textit{Covertness components}, are the other components except malicious and fixed component.
	Based on these three components, an adversary can synthesize a variety of malicious traffic to evade the detection of defenders by imitating benign traffic (such as embedding different instructions in the malicious components). 
	
	In this paper, we focus on the scene that an HTTP-based malware attack the victim devices and we need find a method to detect it during the communication phase. 
	Therefore, we build a hybrid neural network to extract features of malicious traffic, and design a traffic generation algorithm based on the four characteristics of HTTP-based malicious communication traffic. 

\section{Feature Analysis of HTTP-based Malicious Communication Traffic}
We define a flow as a sample.
Flow is full-duplex in application layer, including request and response packets with the same quintuple ($ src\_ip $, $ src\_port $, $ dst\_ip $, $ dst\_port $, $ TCP $) over a period of time. 
A flow has multiple packets. Each packet is composed of payload and different fields, which has hierarchical structures.
Therefore, we divide a flow into packet level (Pkt-level) and flow level (Flow-level) to extract features from different hierarchies. 

\begin{figure}[htbp]
	\vspace{-1.5em}
	\centerline{\includegraphics[scale=1]{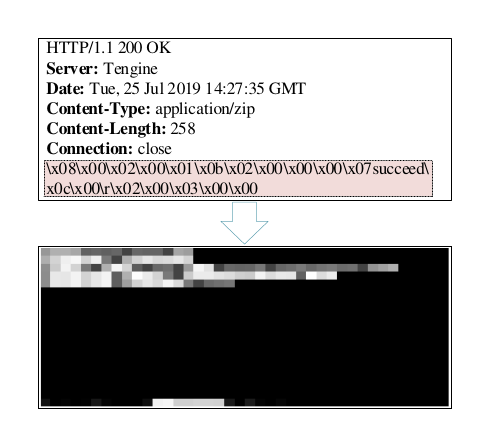}}
	\vspace{-1em}
	\caption{A example of converting a packet(top) of a flow into a two-dimensional image(bottom).}
	\label{pac-demo-gray}
	\vspace{-1.0em}
\end{figure}

\subsection{Feature Analysis of Packet}

\textbf{Text Feature:}
The text content of an HTTP packet is composed of a payload and different header fields.
The feature information of HTTP-based malicious communication behavior is usually hidden in these contents.
For detecting unknown malicious communication behavior, we consider that the miss-leading contents (ip, url, \textit{etc.}), will bring false positives. 
For instance, the host field of the packet can be changed to evade detection when the malicious communication information embedded in a packet is consistent.
Therefore, we drop these miss-leading contents.
Then, a packet is processed as a two-dimensional image with one channel, as shown in Fig.~\ref{pac-demo-gray}.

\textbf{Statistical Features:}
Malicious and benign packets are different in many statistical values.
For instance, an adversary usually uses fewer fields for simplicity. 
The length of the domain name used by the adversary is longer and unreadable due to the occupation of the domain name space. 
Therefore, we propose statistical feature engineering combined with the raw data, to further improve the performance of the detection model. 
The statistical features at Pkt-level are shown in Tab~\ref{PL-vector}. 
RFC1998 \cite{chen1996rfc1998} recommends that web services use 47 field lines for HTTP-based communication, but most web services do not use that much.
Therefore, we consider counting the features of 18 fields. 
It will be discarded if exceeded and be filled with 0 if missed. 

	\begin{table}[htbp]
		\caption{Statistical features at Pkt-level} 
		\begin{center}
				\begin{tabular}{|c|c|}
					\hline
					Type &  Position \\ 
					\hline
					\hline
					packet type & 0 \\
					length of url or state description & 1\\
					protocol version & 2 \\
					lines of fields & 3 \\
					lengths of fields name & 4-21\\
					lengths of fields value & 22-39\\
					length of payload & 40 \\
					\hline
				\end{tabular}
			\label{PL-vector}
		\end{center}
	\end{table}

\subsection{Feature Analysis of Flow}
HTTP-based malicious communication traffic consists of multiple packets in a flow. 
The features provided by a single packet are limited, and analysis based on the entire flow can get more information.
Generally, there are more packets in a malicious flow compared with benign traffic behavior. Usually, the bytes of response are larger but the bytes of request are smaller. 
Therefore, we perform feature analysis at Flow-level. 
Statistics such as the number of packets and length sequences are used as part of statistical feature engineering, as shown in Tab~\ref{FL-vector}. The number of packets of a flow will not exceed 50, if it is exceeded, it will be discarded, or if it is missing, it will be filled with 0. 

	\begin{table}[hbtp]
		\caption{Statistical features at Flow-level}
		\vspace{-1em}
			\begin{center}
				\begin{tabular}{|c|c|}
					\hline
					Type &  Position \\ 
					\hline
					\hline%
					count of request pkts & 0 \\
					count of 'get' & 1 \\
					count of 'post' & 2 \\
					count of 'head' & 3 \\
					count of 'options' & 4 \\
					count of other requests & 5 \\
					count of response pkts & 6 \\
					count of '1XX' & 7 \\
					count of '2XX' & 8 \\
					count of '3XX' & 9 \\
					count of '4XX' & 10 \\
					count of '5XX' and others & 11 \\
					count of other responses & 12 \\
					mean of pkt bytes & 13 \\
					seq of pkts bytes & 14-63 \\
					\hline
				\end{tabular}
				\label{FL-vector}
			\end{center}
	\end{table}


\section{HMCD-Model Methodology}
\begin{figure*}[ht]
	\centerline{\includegraphics[scale=0.51]{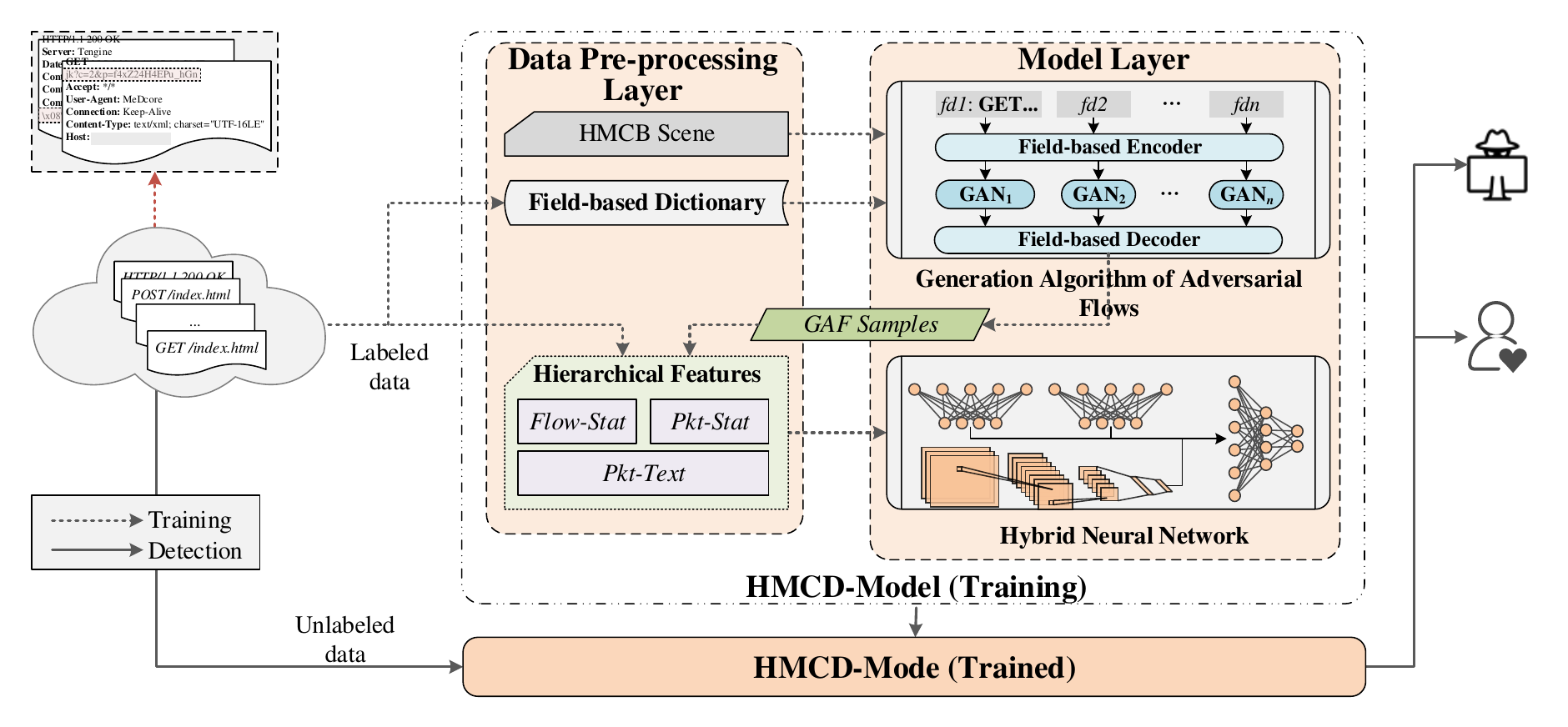}}
	\caption{Training and detection process for HTTP-based malicious communication traffic (HMCB Scene: HTTP-based malicious communication behavior scene).}
	\label{detect-process}
\end{figure*}

\begin{figure}[]
	\vspace{-0em}
	\centerline{\includegraphics[scale=0.53]{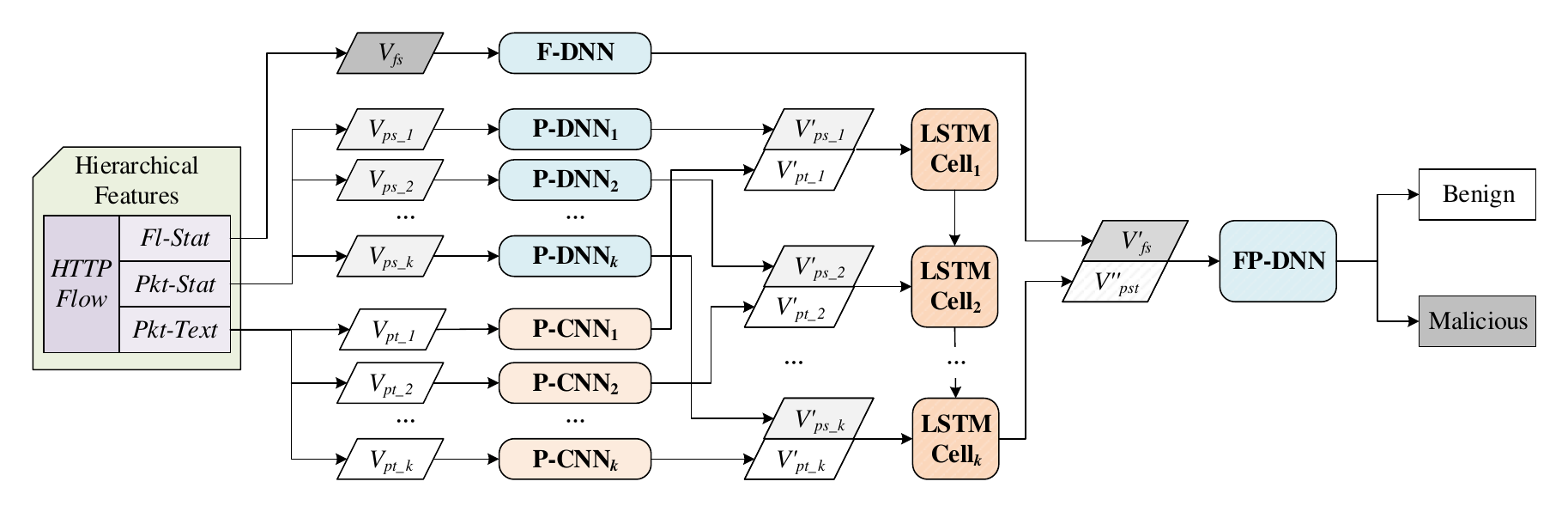}}
	\vspace{-0.5em}
	\caption{Structure of the hybrid neural network ($ V_{fs} $: normalized Flow-Stat vector; $ V_{ps\_*}$: normalized Pkt-Stat vector; $ V_{pt\_*} $: normalized Pkt-Text vector; $ V^{'}_{*} $, $ V^{''}_{*} $: intermediate variables; $ k $: the packet size in an HTTP flow).} 
	\label{model}
	\vspace{-1em}
\end{figure}

\subsection{Overview}

	As shown in Fig.~\ref{detect-process}, we build HMCD-Model. 
	The training and detection process of the model is composed of two different stages.

	\textbf{Generation of Adversarial Flows:} Generate HTTP-based malicious communication traffic for data enhancement. 
	Based on a field-based dictionary, we use WGAN-GP to insert malicious content into benign traffic packets to generate malicious flows, that is Generated Adversarial Flows (GAFs), which are used to enrich the multiformity of training samples.

	\textbf{Training and Detection:} Build a hybrid neural network.
	First, the labeled traffic data from the real world is used for model training, and GAFs are also added to the training set to improve the generalization of the model.
	Then, the trained HMCD-Model fixes the parameters, and performs unknown HTTP-based malicious communication behavior detection in various experimental scenarios. 

Intuitively, we first introduce the hybrid neural network used to extract features, and then introduce the adversarial flows generation algorithm. 

\subsection{Hybrid Neural Network based on Hierarchical Spatial-temporal Features}
The hybrid neural network based on hierarchical spatial-temporal features is shown in Fig.~\ref{model}, and the basic unit of its detection is flow. 
In the spatial, the text features of the packet have structural information. 
So we use CNN to extract content and structural features. 
In the temporal, a flow usually has multiple packets, and there is natural timing between the packets. So we use LSTM to extract its sequence features. 

\subsubsection{Feature Extract at Pkt-level}
A packet has content and structure information. 
We cut a raw packet into a two-dimensional image with one channel to form text feature of the packet. 
In addition, we extract packet statistics to obtain more comprehensive packet information.

\textbf{Text Feature Extract of Packet:}
The hybrid neural network uses CNN to process packet text information (Pkt-Text). CNN exploits convolution and pooling operations, which can perform translation-invariant classification of input information according to its hierarchical structure \cite{Neocognitron}. 
The single element output of the convolution layer corresponds to a matrix feature map input, as shown in Eq~\eqref{conv}, where $ Z^{l} $ is input feature, $ w^{l+1} $ is convolution kernel, $ f_{conv} $ is the size of kernel and $ s_0 $ is the stride. 
Convolution kernel parameters are shared, and the connection between layers are sparse, which allow it to effectively extract features with a small amount of calculation.

The pooling layer performs feature selection and information filtering in the input features. 
As shown in Eq~\eqref{pool}, where $ f_{pool} $ is the size of pooling. 
In this paper, we use maximum pooling ($ p\rightarrow +\infty $) to retain the most significant features. 

\begin{equation}
	\begin{aligned}
		Z^{l+1}(i,j)&=[Z^{l} \otimes w^{l+1} ](i,j)+b \\
		&= \sum_{k=1}^{K_l} \sum_{x, y=1}^{f_{conv} } \left [ Z_{k}^{l} (s_0i+x,s_0j+y)w_{k}^{l+1}(x,y)) \right ]+b \\
		\label{conv}
	\end{aligned}
\end{equation}
\begin{equation}
	\begin{aligned}
		Z_{k}^{l}(i,j)&=\left [ \sum_{x,y=1}^{f_{pool}} Z_{k}^{l}(i+x,j+y)^{p}\right ]^{\frac{1}{p}}
		\label{pool}
	\end{aligned}
\end{equation}

\textbf{Statistical Feature Extract of Packet:}
The statistics of a single packet is a 41-dimensional vector (Pkt-Stat), and there is no obvious characteristic relationship between the components. 
Therefore, we exploit DNN to process the Pkt-Stat and integrating them into lower latitude feature space.

\subsubsection{Feature Extract at Flow-level}
Packets of a flow have natural timing relationship according to the transmission timing. 
These packets form sequence features with a consistent format after packets are processed by CNN and DNN. Therefore, we extract sequence and statistical features in a flow at Flow-level.

\textbf{Sequence Feature Extract of Flow:}
Packets of the same flow are serially transmitted on a timeline to form sequence data, which are suitable for processing using RNN. 
And the packets in the flow may have a long dependency due to delays and re-transmissions. 
LSTM can alleviate the long-term dependency problem \cite{Hochreiter}.
Therefore, we use LSTM instead of traditional RNN to extract the sequence features of flow.
It uses gate structure to selectively remember and forget information. 
Each gate is composed of an activation layer and a pointwise operation. 
By choosing to store information for subsequent processing, information can be transmitted further along the timing chain. 
Sundermeyer \emph{et al.} \cite{sundermeyer2012lstm} show that the most important component in LSTM is the forget gate. 
As shown in Eq~\eqref{forget}, the forget gate, $ f_t $, decides the proportion of the retained information according to the stored information $ h_{t-1} $ and the input $ x_t $ at the current stage.
Followed is the input gate, $ f_t $. And it decides how much new information to add, as shown in Eq~\eqref{input}.

\begin{equation}
	\begin{aligned}
		f_{t} &= \delta (W_{f}\cdot \left [h_{t-1}, x_{t}\right ]+b_{f})
		\label{forget}
	\end{aligned}
\end{equation}
\begin{equation}
	\begin{aligned}
		i_{t} &= \delta(W_{i}\cdot \left[h_{t-1}, x_{t}\right]+b_{i})
		\label{input}
	\end{aligned}
\end{equation}

\textbf{Statistical Feature Extract of Flow:} The statistics of a flow (Flow-Stat) is a 64-dimensional vector. 
There is no obvious relationship between the components. 
Therefore, we exploit DNN to process the vector. 
The input features are dimensional reduced and integrated into low-latitude feature vectors, which combined with LSTM output and then entered into subsequent neurons.

\subsubsection{Feature Aggregation}
FP-DNN takes the sequence and statistical features as input, which processed by LSTM and DNN. 
It exploits fully-connected layers for processing and outputs the final detection result. 
The final detection result of HMCD-Model is output after feature reduction and nonlinear transformation.

\subsubsection{Time Complexity of Hybrid Neural Network}

Time complexity is the number of executions of each operation in an algorithm or model, and is called operation frequency or time frequency. 
The time complexity of the neural network is the number of operations for each basic operation of the model in training and testing. 

The time complexity, $ T_{hnn} $, consists of the time of each local networks during training and testing, including feature extract and feature aggregation, as shown in Eq~\eqref{time-total1}, where $ T_{dnn} $ is the sum of $ T_{p-dnn}$, $ T_{f-dnn} $ and $ T_{fp-dnn} $, $ N $ is the sample size.
When model structure (the size of packet, \textit{etc.}) are determined, its time complexity is linearly related to the sample size, that is, $ O(N) $. 
By the way, in this hybrid neural network, the number of neurons in each component is small to to improve the training and detection speed (convolution kernel, \textit{etc.}). And some local networks can work simultaneously (\textit{e.g.} text feature and statistical feature extract of packet, sequence feature and statistical feature extract of flow), and most neural networks are matrix operations, so there are many parallel computing strategies that can further accelerate the model.

\begin{equation}
	\begin{aligned}
		\begin{split}
			T_{hnn} &= N \cdot \left ( T_{cnn} + T_{lstm}+T_{dnn} \right )\\
			&\sim O\left(\beta \cdot N\right) \sim O\left( N\right)
			\label{time-total1}
		\end{split}
	\end{aligned}
\end{equation}

\subsection{Generation Algorithm of Adversarial Flows}
	In an adversarial environment, the HTTP-based malicious communication traffic collected from Internet is difficult to cover all the specific forms.
	Therefore, we design a generation algorithm for generating adversarial flows.
	The algorithm first builds a dictionary of packet fields (field-based dictionary) based on the real traffic dataset, and then, generates corresponding packets, and combines the packets into flow.
	
	In this paper, each of our generated adversarial flow consists of a request packet and a response packet.
	The process of generation algorithm is described in detail below.
	

\begin{figure}[htbp]
	\centerline{\includegraphics[scale=0.65]{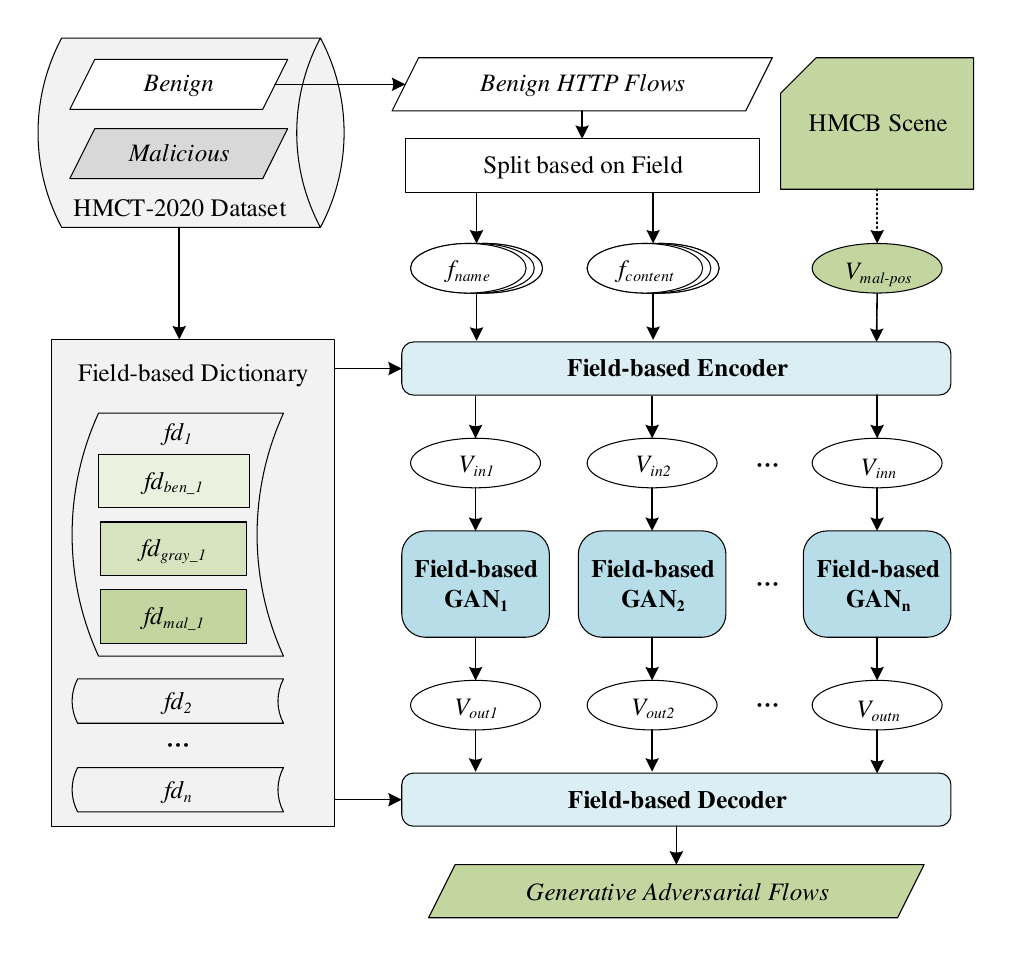}}
	\caption{Process of generating HTTP-based malicious communication traffic (HMCB Scene: HTTP-based malicious communication behavior scene).}
	\label{gans}
\end{figure}

\subsubsection{Generation Process of GAFs}
	The adversary always hides the malicious content in certain packet fields in the scene of HTTP-based malicious communication behavior.
	Therefore, we first generate each content in a flow based on the fields of a packet.
	Then, we splice these field-based contents to synthesize malicious packets, and then splice these packets to flows.
	
	The process of generating HTTP-based malicious communication traffic is shown in Fig.~\ref{gans}. The process is divided into: 1) We analyze the traffic and generate corresponding benign and malicious dictionaries. 2) We randomly select the malicious content in the malicious dictionary and determine the location where it appears in HTTP benign flows, and finally obtain the corresponding field code $ V_{in} $. 3) We input multiple $ V_{in} $ of a packet to the corresponding GAN for training, and output the generated field code $ V_{out} $. 4) According to the HTTP specification, multiple $ V_{out} $ are decoded to packets and spliced into complete malicious communication flow.

\textbf{Field-based Dictionary:}
	We build a field-based dictionary, $ field\_dict $, as shown in Eq~\eqref{dict1}. $ field\_dict $ is a two-level dictionary and contains three second-level dictionaries (malicious dictionary ($ fd_{mal} $,) gray dictionary ($ fd_{gray} $) and benign dictionary ($ fd_{ben} $)).
	HTTP packets have fixed structure, each row is independent and has specific meaning. 
	Therefore, we divide packets by row, and each row is split into a fixed field-based name $ f_{name} $ (“GET”, “Accept”, \textit{etc.}) and a remaining fixed field-based content $ f_{content} $ (may contain malicious information). 
	The key of the first-level dictionary is $ f_{name} $ (“Date”, \textit{etc.}) and the value is a second-level dictionary. 
	According to the syntax and semantics of the $ f_{content} $, we divide it into several words (“text”, “xml”, \textit{etc.}) by special characters (“,”, “:”, \textit{etc.}).

\begin{equation}
	\begin{aligned}
		field\_dict = \begin{Bmatrix}
			f_{name\_1}: fd_{mal\_1}+fd_{gray\_1}+fd_{ben\_1}\\ 
			f_{name\_2}: fd_{mal\_2}+fd_{gray\_2}+fd_{ben\_2}\\ 
			...\\ 
			f_{name\_n}: fd_{mal\_n}+fd_{gray\_n}+fd_{ben\_n}
		\end{Bmatrix}
		\label{dict1}
	\end{aligned}
\end{equation}

For instance, we show an example in Fig.~\ref{attack-demo}. These two packets are divided into 11 pairs of $ f_{names} $, and $ f_{contents} $. 
In the first one, $ f_{name} $ is “GET”, and $ f_{content} $ (“jk?c=2\&p=f4Z24…”) can be split into words (“jk”, “c”, “2”, “p”, “f4Z24...”, $ etc. $) by the characters like “?”, “=”, “\&”. 
We use words that appear more than $ p $ times as keys to make the dictionary statistically stable and control the size of dictionary by adjusting the value of $ p $. 
In addition, keys of $ fd_{mal} $ only appear in malicious HTTP packets, keys of $ fd_{ben} $ only appear in benign HTTP packets, and keys of $ fd_{gray} $ appear in both.

\textbf{Encoder:}
According to the analysis about HTTP-based malicious communication behavior (see Section 3), we set the vector $ V_{mal-pos} $ to present the possible positions of packets in a flow where adversaries usually hide malicious content.
As shown in Eq~\eqref{mal-loca}, where $ p_i $ is the replacement position, $ w_i $ is the malicious content word randomly selected from the keys in the $ fd_{mal} $, $ m $ is length of the $ V_{mal-pos} $.
Several parts of benign contents are replaced with malicious contents based on $ V_{mal-pos} $. Then, the field-based contents through replacement are encoded as integer vectors ($ V_{in\_1} $, $ V_{in\_2} $, \textit{etc.}). 

\begin{equation}
	\begin{aligned}
		V_{mal-pos} = \left \{ \left \langle p_1,w_1 \right \rangle,\left \langle p_2,w_2 \right \rangle,...,\left \langle p_m,w_m \right \rangle  \right \}
		\label{mal-loca}
	\end{aligned}
\end{equation}

\begin{figure}[htbp]
	\vspace{-1.5em}
	\centerline{\includegraphics[scale=0.75]{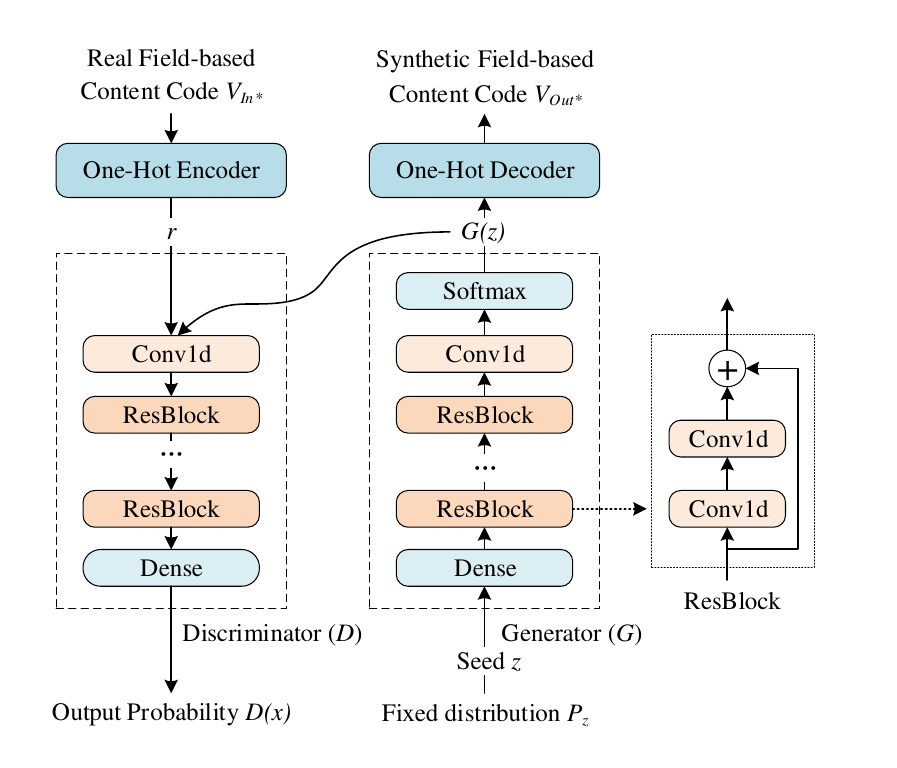}}
	\vspace{-1em}
	\caption{Structure of Field-based GAN.}
	\label{gan}
	\vspace{-0.5em}
\end{figure}

\textbf{Field-based GAN:} The Field-based GAN is responsible to generate malicious content which is highly similar to benign content. 
As shown in Fig.~\ref{gan}, Field-based GAN consists of two modules: the Generator ($ G $) and the Discriminator ($ D $). The structure of the $ G $ is as same as $ D $, but their order is reversed. 
$ G $ is to imitate real field-based benign contents and generate fake, and $ D $ is to judge whether the generated contents by $ G $ are similar to the real benign contents.

In Field-based GAN, we introduce the idea of WGAN-GP \cite{gulrajani2017improved}, which makes Field-based GAN to generate diverse data. WGAN-GP defines a well-defined training procedure, which makes the generator easier to train than traditional GAN.
It describes the distance between $ p_g $ and $ p_r $ by the Wasserstein distance.
WGAN-GP can be considered a min-max game, as shown in Eq~\eqref{wgan-gp}, where $ z $ is the input of $ G $ which is a random 1-D tensor $ z $ filled, $ R $ is the regular term and $ \mathbb{E} $ is the cross-entropy function.
Through Eq~\eqref{wgan-gp}, Field-based GAN continuously reverses iterative neural network training $ G $ and $ D $, takes the random seed $ z $ as the input of $ G $, output the content sample $ G(z) $, and takes the $ V_{in} $ and $ G(z) $ (one-hot form) as the input of $ D $.
When all parameters converge, we decode the generated $ G(z) $ into the generated content $ V_{out} $.

\begin{equation}
	\begin{aligned}
		\underset{G}{min}\: \underset{D}{max}\: L(D,G)=\mathbb{E}_{z\sim P_z}\left [ D(G(z)) \right ] - \mathbb{E}_{x\sim P_r}\left [ D(x) \right ] + R 
		\label{wgan-gp}
	\end{aligned}
\end{equation}


\textbf{Algorithm Description:} The generation algorithm of adversarial flows based on the above-described generation process is shown in Algorithm~\ref{alg2}. 
The input of the algorithm is the $ f_{name} $ and $ f_{content} $ of HTTP benign flows and the $ V_{mal-pos} $. The output is GAFs. 
Algorithm guarantees \textit{maliciousness} of GAFs through $ V_{mal-pos} $. 
The strategy of dividing packets based on fields ensures \textit{compliance} of the generated traffic. Field-based GAN will imitate the characteristics of benign traffic when generating traffic, which ensures \textit{covertness} of the generated traffic. 
In addition, we use the WGAN-GP to optimize the training process and enhance \textit{multiformity} of GAFs.

\begin{algorithm} 
	\caption{generation algorithm of adversarial flows} 
	\label{alg2} 
	\begin{algorithmic}[19] 
		\REQUIRE  \textit{benign-flows} : HTTP-based benign traffic. $ V_{mal-pos} $: the vector that may contain malicious information. $ N $ : the number of samples that need to generate.
		\ENSURE $ g\_flows $: generated HTTP-based malicious communication flows.
		\STATE{\textbf{Step 1:} Dictionaries}
		\STATE{Segement words and build dictionaries $ field\_dict $ from HMCD-2020 datasets}
		\STATE{\textbf{Step 2:} Encoder}
		\FOR{ $ f_{field} $ in $field\_dict $} 
		\STATE{ The content-string is separated from the corresponding fields of benign traffic
		}
		\STATE{	Encode the content-string as $ V_{code} $ based on $ field\_dict $
		}
		\ENDFOR 
		\STATE {\textbf{Step 3:} Field-basd GAN}
		\FOR{t=0 in number of iterations}
		\STATE{The $ seed (z) $ is transformed into the sample $ G(z) $ with the same dimension as $ V_{code} $}
		\STATE {Convert $ G(z) $ to $ V_{mal-code} $ based on $ V_{mal-pos} $}
		\STATE {Take $ V_{code} $ and $ V_{mal-code} $ as inputs to $ D $}
		\STATE{Update $D$ model and $G$ model by Eq~\eqref{wgan-gp}}
		\ENDFOR
		\STATE {\textbf{Step 4:} Generated Adversarial Flows (GAFs)}
		\STATE{Decode the generated $ G(z) $ and merge it into a malicious communication flow $ g\_flow $}
		\STATE{$ g\_flows $ = $ N $ generated HTTP-based malicious communication flows}
	\end{algorithmic} 
\end{algorithm}

\subsubsection{Time Complexity of Generation Algorithm}


The algorithm traverses the input data once in building the dictionary and encoding stage. In the generation stage, the time complexity ($ T_{ga} $) is shown in Eq~\eqref{gancom}, where $ N $ is the sample size, $ F $ is the number of Field-based GAN, $ C_l $ is the number of convolution kernels of l-th layer, $ M $ is the size of the feature map, and $ K $ is kernel size, $ D_{in} $ and $ D_{out} $ are the input and output of the fully-connected layer. 
All the parameters are constants after the model structure is determined. Therefore, the time complexity of the entire generation algorithm ($ T_{ga} $) can be regarded as $ O(N) $.

\begin{equation}
	\begin{aligned}
		\begin{split}
			T_{ga} &\sim O\left ( N \cdot F \cdot \sum_{l=1}^{L} \left ( M_{l}^{2}\cdot K_{l}^{2} \cdot C_{l-1}\cdot C_{l} + 2 \cdot D_{in} \cdot D_{out}\right ) \right ) \\
			&\sim O\left(\alpha \cdot N\right) \sim O\left( N\right)
			\label{gancom}
		\end{split}
	\end{aligned}
\end{equation}

\section{Experimental Evaluation}
\newcommand{\tabincell}[2]{\begin{tabular}{@{}#1@{}}#2\end{tabular}}

\subsection{Datasets}
\subsubsection{HMCT-2020 Dataset}

	Currently, there are no large-scale public intrusion detection datasets specifically for HTTP-based malicious communication behavior detection, the relevant datasets that contain HTTP-based malicious communication traffic have limited specific attack forms. 
	Therefore, we publish the dataset HMCT-2020 and the hashing technology is used for data masking to protect privacy.
	HMCT-2020 can be divided into two datasets from the timeline, HMCT-2020(18) and HMCT-2020(19-20).
	The traffic in HMCT-2020(18) is mainly captured from the Internet during November 2018. The traffic in HMCT-2020(19-20) is captured from the Internet between July 2019 and March 2020.
	Finally, after data cleaning and application layer packet extraction, the captured traffic forms the dataset HMCD-2020, the details of which are shown in Tab~\ref{tab2}. The dataset can be found in there\footnote{https://github.com/BitBrave-Xie/HMCD-Model}.
	

	\begin{table}[ht]
		\caption{Statistics on packet size and flow size in HMCT-2020}
		\begin{center}
			\setlength{\tabcolsep}{0.4mm}{
				\begin{tabular}{|c|c|c|c|c|c|c|c|c|c|}
					\hline
					& \multicolumn{4}{c|}{Packet (in bytes)} & & \multicolumn{4}{c|}{Flow (in packtes)} \\
					\cline{2-5}
					\cline{7-10}
					& \diagbox{\ \ \ \ \ \ \ \ }{\ } & Malicious & Benign & Total & & \diagbox{\ \ \ \ \ \  }{\ } & Malicious & Benign & Total \\
					\hline
					\multirow{5}{*}{\tabincell{c}{HMCT-2020 \\ (18)}}  & \tabincell{c}{Count \\(in packets)} & 1.49M & 11.48M & 12.97M & & \tabincell{c}{Count \\(in flows)}  & 35.58K & 3.92M & 3.95M \\
					\cline{2-5}
					\cline{7-10}
					& Size & 429.96M & 6.5G & 6.93G & & Size & 1.49M & 11.48M & 12.97M \\
					\cline{2-5}
					\cline{7-10}
					& Min & 15 & 12 & 12 & & Min & 2 & 1 & 1 \\
					\cline{2-5}
					\cline{7-10}
					& Max & 46.55K & 10.23K & 46.55K & & Max & 50 & 50 & 50 \\
					\cline{2-5}
					\cline{7-10}
					& Mean & 289.25 & 566.33 & 534.58 & & Mean & 41.77 & 2.93 & 3.28 \\
					\hline
					\multirow{5}{*}{\tabincell{c}{HMCT-2020 \\ (19-20)}}  & \tabincell{c}{Count \\(in packets)} & 68.77K & 2.86M & 2.93M & & \tabincell{c}{Count \\(in flows)} & 41.18K & 882.0K & 923.17K \\
					\cline{2-5}
					\cline{7-10}
					& Size & 29.6M & 1.73G & 1.76G & & Size & 68.77K & 2.86M & 2.93M \\
					\cline{2-5}
					\cline{7-10}
					& Min & 15 & 19 & 15 & & Min & 1 & 1 & 1 \\
					\cline{2-5}
					\cline{7-10}
					& Max & 6.32K & 7.08K & 7.08K & & Max & 50 & 50 & 50 \\
					\cline{2-5}
					\cline{7-10}
					& Mean & 430.41 & 604.43 & 600.34 & & Mean & 1.67 & 3.24 & 3.17 \\
					\hline
			\end{tabular}}
			\label{tab2}
		\end{center}
	\end{table}

	There are two sources of malicious traffic in HMCT-2020. One is to directly extract relevant malicious traffic from the network operator's existing IDSs. The other is to actively capture traffic by running target malware in a sandbox. In HMCT-2020(18), a total of 35,583 flows are obtained, and in HMCT-2020(19-20), a total of 41,177 flows are obtained. 

	Benign traffic comes from the network gateway of our network security lab. After authorization, we deploy a traffic collection device at the gateway to collect traffic while ensuring security and data privacy. Finally, HMCT-2020(18) contains the benign flows of 3,981,567, and HMCT-2020(19-20) contains the benign flows of 923,172. 

	We employ multiple ways to ensure the validity of traffic data labels in HMCT-2020.
	First, malicious traffic and benign traffic originate from different channels, and are naturally distinguishable so that they can be self-labeled.
	For instance, we class malicious traffic captured by partner operators' existing IDSs and traffic generated by malware running in sandboxes as malicious, and traffic captured from lab network gateways as benign. 
	Second, after cleaning the traffic data, we randomly sample from the preliminary labeled data to make artificial judgments to further increase the accuracy of data labeling.
	Finally, we construct the labeled dataset HMCT-2020.

\subsubsection{Other Datasets}
In this paper, the malicious traffic in the published intrusion detection dataset and the malicious traffic generated by the malware are collected to further test the generalization of model.

\textbf{CIC-IDS-2017:}
CIC-IDS-2017\cite{sharafaldin2018toward} is an intrusion detection dataset with complete traffic published by the Canadian Institute for Cybersecurity at University Of New Brunswick, in 2018. It contains benign traffic and latest common attacks (Web Attack, Infiltration, \textit{etc.}). We select all HTTP-based malicious traffic as positive samples and all the background traffic as the unknown benign traffic. The specific dataset details are shown in Tab~\ref{tab3}. 

\begin{table}[ht]
	\caption{Statistics on packet size and flow size in CIC-IDS-2017}
	\begin{center}
		\setlength{\tabcolsep}{0.4mm}{
			\begin{tabular}{|c|c|c|c|c|c|c|c|c|}
				\hline
				\multicolumn{4}{|c|}{Packet (in bytes)} & & \multicolumn{4}{c|}{Flow (in packets)} \\
				\cline{1-4}
				\cline{6-9}
				\diagbox{ \ \ \ \ \ \ \ }{ \ \ } & Malicious & Benign & Total & & \diagbox{ \ \ \ \ \ \  }{\ } & Malicious & Benign & Total \\
				\hline
				  \tabincell{c}{Count \\(in packets)} & 131.1K & 438.58K & 569.68K & & \tabincell{c}{Count \\(in flows)} & 27.47K & 106.23K & 133.7K \\
				\cline{1-4}
				\cline{6-9}
				Size & 285.21M & 580.78M & 865.99M & & Size & 131.1K & 438.58K & 569.68K \\
				\cline{1-4}
				\cline{6-9}
				Min & 16 & 16 & 16 & & Min & 2 & 1 & 1 \\
				\cline{1-4}
				\cline{6-9}
				Max & 20.17K & 24.71K & 24.71K & & Max & 50 & 50 & 50 \\
				\cline{1-4}
				\cline{6-9}
				Mean & 2.18K & 1.32K & 1.52K & & Mean & 4.77 & 4.13 & 4.26 \\
				\hline
		\end{tabular}}
		\label{tab3}
	\end{center}
\end{table}

\textbf{82-Malware-Traffic:} We collect HTTP traffic generated by 82 malware from an influential malware web\footnote{http://www.malware-traffic-analysis.net/index.html}. The selection condition is malware with behavioral characteristics that meet the conditions of HTTP-based malicious communication behavior, such as \textit{Gootkit}. In addition, we collect a large amount of background traffic from different Internet gateways as unknown benign traffic. The details about traffic can be found in our publish link (in Introduction). The specific dataset details are shown in Tab~\ref{tab0}. 

\begin{table}[ht]
	\caption{Statistics on packet size and flow size in 82-Malware-Traffic}
	\begin{center}
		\begin{tabular}{|c|c|c|c|c|}
			\hline
			\multicolumn{2}{|c|}{Packet (in bytes)} & & \multicolumn{2}{c|}{Flow (in packets)} \\
			\cline{1-2}
			\cline{4-5}
			\diagbox{\ \ \ \ \ \ \ }{\ } & Malicious & & \diagbox{\ \ \ \ \ \ \ }{\ } & Malicious \\
			\hline
			\tabincell{c}{Count \\(in packets)} & 70.19K & & \tabincell{c}{Count \\(in flows)} & 3.19K \\
			\cline{1-2}
			\cline{4-5}
			Size & 22.94M & & Size & 70.19K \\
			\cline{1-2}
			\cline{4-5}
			Min & 15 & & Min & 1 \\
			\cline{1-2}
			\cline{4-5}
			Max & 7.41K & & Max & 50 \\
			\cline{1-2}
			\cline{4-5}
			Mean & 326.81 & & Mean & 21.98 \\
			\hline
		\end{tabular}
		\label{tab0}
	\end{center}
\end{table}

\subsection{Evaluation Metrics and Environmental Configuration}
\subsubsection{Evaluation Metrics}
	There are 4 basic metrics in the experiment. True-Positive ($ TP $), is the number of malicious samples classified as malicious.
	False-Positive ($ TP $), is the number of benign samples classified as malicious.
	True-Negative ($ TN $), is the number of benign samples classified as benign.
	False-Negative ($ FN $), is the number of malicious samples classified as benign.
	
	Based on the above metrics, We use precision ($P$), recall ($R$) and false positive rate ($FPR$) to evaluate a model's performance in detecting HTTP-based malicious communication behavior, as shown in Eq\eqref{eq1} to Eq\eqref{eq3}.
	In addition, $F1$ is also used to verify the overall performance of a model in detecting malicious behavior.
	As shown in Eq\eqref{eq4}, $ P_{k} $ and $ R_{k} $ are the precision and recall of the model in the k-th experiment.
	Because the same experiment will be repeated $ N_r $ times, it will produce multiple indicators of the same type, here we calculate the macro average to get the final result.

\begin{equation}
	\begin{aligned}
		P &= \frac{1}{N_r}\sum_{k=1}^{N_r}\frac{TP_{k}}{TP_{k}+FP_{k}}\\ 
	\end{aligned}\label{eq1}
\end{equation}
\begin{equation}
	\begin{aligned}
		R &= \frac{1}{N_r}\sum_{k=1}^{N_r}\frac{TP_{k}}{TP_{k}+FN_{k}}\\
	\end{aligned}\label{eq2}
\end{equation}
\begin{equation}
	\begin{aligned}
		FPR &= \frac{1}{N_r}\sum_{k=1}^{N_r}\frac{FP_{k}}{FP_{k}+TN_{k}}\\
	\end{aligned}\label{eq3}
\end{equation}

\begin{equation}
	\begin{aligned}
		F1&= \frac{1}{N_r}\sum_{k=1}^{N_r}\frac{2\times P_{k} \times R_{k}}{P_{k}+R_{k}}\\
	\end{aligned}\label{eq4}
\end{equation}

\subsubsection{Environmental Configuration}
	The system environment is Ubuntu16.04 LTS. The hardware facilities are 16-core CPU and 128G memory. TensorFlow2.0 in Python3.7 is used to implement the model. To accelerate training and detection, 3 NVIDIA TITAN XPs are deployed on the server.

	As a rule of thumb, we set some hyper-parameters for the model and related experiments. Field-based GAN has 7 convolutional layers and 2 full-connected layers. CNN has one convolutional layer with two $2 \times 8$ neurons. LSTM cell uses 16 neurons. The structure of DNN is 10, 8, 2. The other parameters about HMCD-Model are shown in Tab~\ref{detail}. 
	
	\begin{table}[htbp]
		\caption{Parameter configuration of HMCD-Model during training}
		\begin{center}
				\begin{tabular}{|c|c|}
					\hline
					Type &  Value \\ 
					\hline
					\hline
					Loss function & Cross-Entropy \\
					Activation function & ReLU \\
					Optimizer & Adam\\
					learning rate & 1e-3 \\
					Batch size & 128\\
					Epochs & 50 \\
					\hline
				\end{tabular}
			\label{detail}
		\end{center}
	\end{table}

	In addition, we set a sample consist the first two packets in a flow, and the size of a packet is $ 20\times 40 $. The number of repeated experiments is $ N_r=5 $. During the training of model, the 5-fold cross validation is used.

	
	
	For the experimental data, we conduct experiments by randomly sampling part of the traffic data from the datasets. Specifically, for the training set data, we randomly sample it in HMCT-2020(18), and for the test set data, we randomly sample it from different datasets according to the different requirements of the experiment. Tab~\ref{tab123} shows the details of the data composition of the training and test sets for each experiment.
	For instance, when detecting 82-Malware-Traffic, we select all malicious traffic. 
	In addition, in order to ensure that the experimental results can truly reflect the performance of the model in the entire dataset, all data in each experiment are re-sampled randomly from the corresponding dataset.
	
	
	\begin{table}[hbtp]
		\caption{Data composition of training set and test set in different experiments} 
		\begin{center}
			\setlength{\tabcolsep}{1.5mm}{
				\begin{tabular}{|c|c|c|c|c|c|c|}
					\hline
					\multirow{2}{*}{\diagbox[height=43pt,innerrightsep=10pt]{\ }{\ }} & \multicolumn{3}{c|}{Training set}& \multicolumn{3}{c|}{Test set} \\
					\cline{2-7} 
					& {Data source} & \tabincell{c}{Malicious \\ samples} & \tabincell{c}{Benign \\ samples} & {Data source} & \tabincell{c}{Malicious \\ samples} & \tabincell{c}{Benign \\ samples} \\
					\hline
					{ep1} & HMCT-2020(18) & 20,000 & 50,000 & HMCT-2020(18) & 10,000 & 10,000 \\
					\hline
					{ep2} & HMCT-2020(18) & 20,000 & 50,000 & HMCT-2020(19-20) & 30,000 & 30,000 \\
					\hline
					{ep3} & HMCT-2020(18) & 20,000 & 50,000 & CIC-IDS-2017 & 27,474 & 30,000 \\
					\hline
					{ep4} & HMCT-2020(18) & 20,000 & 50,000 & 82-Malware-Traffic & 3,138 & 4,000 \\
					\hline
				\end{tabular}}
			\label{tab123}
		\end{center}
	\end{table}
	
	For the number of GAFs in the training set, theoretically, the more GAFs, the higher the generalization of the HMCD-Model, but our preliminary experimental results show that it will reduce the performance of the model when the number of GAFs in the training set is too large.
	This is because a generated adversarial flow sample we generate consists of a request and a response, which can only imitate part of the HTTP-based real malicious communication behavior. The generated data can only change in a certain local feature space, so the excessive proportion of GAFs in the training set will lead to the performance of the model overfitting and reduce its generalization performance. 
	In subsequent experiments, in the combination of training and test sets ep1 to ep4, we add 10,000 GAFs to the training set. That is, there are 30,000 malicious samples in the training set, consisting of 20,000 real traffic samples and 10,000 GAFs.


\subsection{Ablation Study on Key Factors in HMCD-Model}


We add statistical features to improve the detection performance of HMCD-Model and GAFs to improve the data multiformity. We verify these key factors in HMCT-2020 and observe corresponding improvements.

\begin{table}[hbtp]
	\caption{Performance improvements of HMCD-Model due to the addition of statistical features and GAFs (ep2 in Tab~\ref{tab123})}
	\begin{center}
			\begin{tabular}{|c|c|c|c|c|}
				\hline
				\multirow{2}{*}{\diagbox[height=35pt,innerrightsep=45pt]{\makecell[l]{\ }}{\ }} & \multicolumn{4}{c|}{\tabincell{c}{HMCT-2020(19-20) \\ (ep2)}} \\
				\cline{2-5} 
				& $ P(\%) $ & $ R(\%) $ & $ F1(\%) $ & $ FPR(\%) $ \\
				\hline
				\tabincell{c}{HMCD-Model without \\ Statistical Features}& {97.33$ \begin{matrix} +0.75\\ -0.62\end{matrix}$} & {99.37$ \begin{matrix} +0.27\\ -0.16\end{matrix}$} & {98.33$ \begin{matrix} +0.34\\ -0.32\end{matrix}$} & {2.73$ \begin{matrix} +0.65\\ -0.79\end{matrix}$} \\
				\hline
				\tabincell{c}{HMCD-Model \\ without GAFs} & {94.63$ \begin{matrix} +3.86\\ -4.50\end{matrix}$} & {95.74$ \begin{matrix} +2.68\\ -5.96\end{matrix}$} & {95.15$ \begin{matrix} +3.31\\ -3.62\end{matrix}$} & {5.50$ \begin{matrix} +5.15\\ -3.99\end{matrix}$} \\
				\hline
				HMCD-Model & {\textbf{97.94}$ \begin{matrix} +0.7\\ -1.51\end{matrix}$} & {\textbf{99.39}$ \begin{matrix} +0.18\\ -0.25\end{matrix}$} & {\textbf{98.66}$ \begin{matrix} +0.38\\ -0.89\end{matrix}$} & {\textbf{2.10}$ \begin{matrix} +1.57\\ -0.73\end{matrix}$} \\
				\hline
			\end{tabular}
		\label{tab45}
	\end{center}
\end{table}

\subsubsection{Statistical Features}
We extract statistical features from Pkt-level and Flow-level. The addition of statistical features can make HMCD-Model get more comprehensive information and easier to extract the essential features of the flow. Traffic in HMCT-2020(18) is used training set, the experimental results are shown in Tab~\ref{tab45}. All the indicators of HMCD-Model have been improved after adding statistical features. 

\subsubsection{Generated Adversarial Flows}
\begin{figure}[htbp]
		\vspace{-1em}
		\centerline{\includegraphics[scale=0.55]{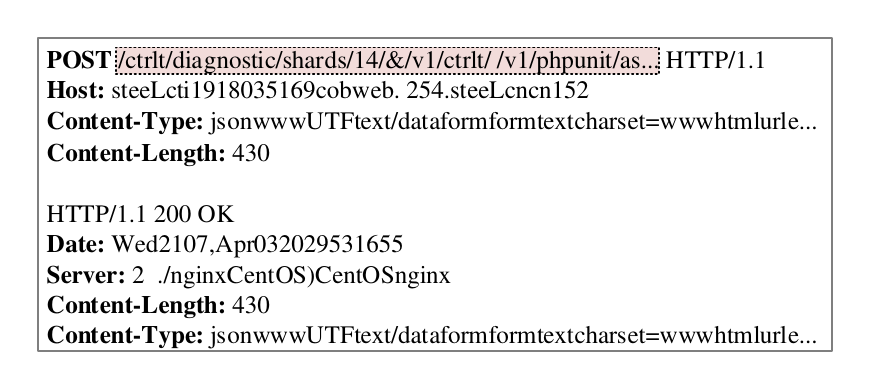}}
		\vspace{-1em}
		\caption{A generated adversarial flow composed of two packets (a request and a response).}
		\label{attack-demoXX}
	\end{figure}

	GAFs are used for data enhancement, to enrich the data multiformity of the training set. The experimental results are shown in Tab~\ref{tab45}. After adding GAFs, the model can be increased to 98.66\% in $ F1 $, and the \textit{FPR} can be reduced to 2.10\%. GAFs expand the training set so that the model can learn more malicious behavior scenes, reduce over-fitting, and improve its generalization ability. We show a generated adversarial flow in  Fig.~\ref{attack-demoXX}, which consists of a request and a response.

\begin{figure}[htbp]
	\centerline{\includegraphics[scale=0.45]{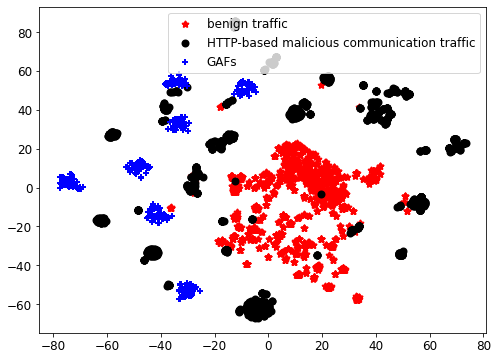}}
	\caption{The feature distribution of HTTP-based malicious communication traffic, benign traffic and GAFs based on t-SNE dimension reduction.}
	\label{gan-tsne}
\end{figure}

	In addition, we visualize the partial sample features distribution from the training set of HMCT-2020(18) and the corresponding GAFs feature distribution.
	The feature distribution based on t-SNE dimension reduction is shown in Fig.~\ref{gan-tsne}.
	HTTP-based malicious communication traffic and benign traffic are distributed in different feature spaces.
	Benign traffic can be grouped into a large cluster as a whole because they usually have some common characteristics.
	HTTP-based malicious communication traffic is distributed in multiple small clusters and the whole is relatively scattered.
	Because the behavior and purpose of different attackers represented by these flows are different, they often construct flows with extremely different characteristics to bypass different detection systems.
	The distribution of GAFs also shows the same phenomenon of small cluster distribution as HTTP-based malicious communication traffic.
	They are distributed in the feature space outside the benign traffic.
	This shows that our generation algorithm can effectively generate adversarial samples and fill the feature space of HTTP-based malicious communication traffic.

\subsection{Generalization Comparison}

	We investigate research work related to HTTP-based malicious communication behavior and selected two well-represented works. Wang \emph{et al.} \cite{wang2017detecting} regard traffic as language and use N-gram and SVM for detection, which can obtain 99.15\% accuracy when detecting malware flows and can detect 54.81\% for unknown malicious applications. Salo \emph{et al.} \cite{salo2019dimensionality} propose a novel hybrid dimensionality reduction technique for intrusion detection combining IG and PCA with an ensemble classifier based on SVM, IBK, and MLP, named IG-PCA-Ensemble. It can achieve 99+\% performance in dataset ISCX-2012, NSL-KDD and Kyoto2006+. 

	We implement these two methods and select hyper-parameters of the two methods according to the trade-off strategy between detection performance and detection time in the experiments. 

	N-gram+SVM: Wang \textit{et al.} We use two packets as the input feature for each flow. For other hyper-parameters, we follow the selection of the original paper under the reasonable time cost. We set the word length $ N =1 $ in N-gram and the feature number $ K=600 $ for each sample according to Wang's experiment results.

	IG-PCA-Ensemble: We set the number of neighbors of IBK to six, SVM is a linear kernel, and MLP has 128 neurons. It reads the first two packets for each flow, selects the top 17 most important information through IG, and reduces to 12-dimensional features based on PCA.

	We compare the generalization of HMCD-Model, N-gram+SVM, and IG-PCA-Ensemble in a variety of experimental environments. In addition, we also compare HMCD-Model with classic baseline machine learning methods in 82-Malware-Traffic (see Appendix A). 




\subsubsection{Generalization Comparison in Dataset HMCT-2020}

We select data from the HMCT-2020(18) to form the training set and the test set. The results of experiments are shown in Tab~\ref{tabXXX2017}. It shows that HMCD-Model has the best detection performance overall. In HMCT-2020(18), compared with N-gram+SVM, HMCD-Model has the same level of detection performance, but its $ F1 $ $\approx $ 99.46\% is 13.8\% higher than IG-PCA-Ensemble (the $ F1 $ $\approx $ 85.66\%).


	\begin{table}[bhtp]
	\caption{Performance comparison in dataset HMCT-2020(18)  (ep1 in Tab~\ref{tab123})}
	\begin{center}
		\setlength{\tabcolsep}{1mm}{
			\begin{tabular}{|c|c|c|c|c|}
				\hline
				\multirow{2}{*}{\diagbox[height=35pt,innerrightsep=40pt]{\makecell[l]{\ }}{\ }} & \multicolumn{4}{c|}{\tabincell{c}{HMCT-2020(18) \\ (ep1)}} \\
				\cline{2-5} 
				& $ P(\%) $ & $ R(\%) $ & $ F1(\%) $ & $ FPR(\%) $ \\
				\hline
				N-gram+SVM & {\textbf{99.79}$ \begin{matrix} +0.11\\ -0.07\end{matrix}$} & {98.92$ \begin{matrix} +0.62\\ -2.18\end{matrix}$} & {99.35$ \begin{matrix} +0.32\\ -1.10\end{matrix}$} & {\textbf{0.21}$ \begin{matrix} +0.07\\ -0.11\end{matrix}$} \\
				\hline
				IG-PCA-Ensemble & {83.92$ \begin{matrix} +4.89\\ -5.73\end{matrix}$} & {87.71$ \begin{matrix} +1.93\\ -2.71\end{matrix}$} & {85.66$ \begin{matrix} +1.31\\ -2.13\end{matrix}$} & {17.21$ \begin{matrix} +7.79\\ -6.50\end{matrix}$} \\
				\hline
				HMCD-Model & {99.52$ \begin{matrix} +0.34\\ -0.20\end{matrix}$} & {\textbf{99.39}$ \begin{matrix} +0.43\\ -0.41\end{matrix}$} & {\textbf{99.46}$ \begin{matrix} +0.19\\ -0.30\end{matrix}$} & {0.48$ \begin{matrix} +0.20\\ -0.34\end{matrix}$} \\
				\hline
		\end{tabular}}
		\label{tabXXX2017}
	\end{center}
	\end{table}

	In addition, we use the data of HMCT-2020(18) for training and the data of HMCT-2020(19-20) with more malicious types for testing to examine the generalization of the model against concept migration. The experimental results of the three methods are shown in Tab~\ref{tab9}. Obviously, HMCD-Model has significantly better detection performance than the other two methods.
	For instance, HMCD-Model's $ FPR $ is only 2.1\%, and the $ F1 $ has a 35.67\% and 26.07\% improvement over N-gram+SVM, IG-PCA-Ensemble, respectively. 

	\begin{table}[htbp]
		\caption{Generalization comparison in dataset HMCT-2020(19-20)  (ep2 in Tab~\ref{tab123})} 
		\begin{center}
				\begin{tabular}{|c|c|c|c|c|}
					\hline
					\multirow{2}{*}{\diagbox[height=35pt,innerrightsep=40pt]{\makecell[l]{\ }}{\ }} & \multicolumn{4}{c|}{\tabincell{c}{HMCT-2020(19-20) \\ (ep2)}} \\
					\cline{2-5} 
					& $ P(\%) $ & $ R(\%) $ & $ F1(\%) $ & $ FPR(\%) $ \\
					\hline
					N-gram+SVM & {81.79$ \begin{matrix} +1.83\\ -1.07\end{matrix}$} & {51.24$ \begin{matrix} +1.31\\ -1.24\end{matrix}$} & {62.99$ \begin{matrix} +0.73\\ -0.84\end{matrix}$} & {11.42$ \begin{matrix} +1.08\\ -1.34\end{matrix}$} \\
					\hline
					IG-PCA-Ensemble & {75.92$ \begin{matrix} +6.05\\ -6.21\end{matrix}$} & {70.07$ \begin{matrix} +4.68\\ -4.87\end{matrix}$} & {72.59$ \begin{matrix} +0.60\\ -0.45\end{matrix}$} & {23.02$ \begin{matrix} +9.46\\ -8.68\end{matrix}$} \\
					\hline
					HMCD-Model & {\textbf{97.94}$ \begin{matrix} +0.70\\ -1.51\end{matrix}$} & {\textbf{99.39}$ \begin{matrix} +0.18\\ -0.25\end{matrix}$} & \textbf{{98.66}$ \begin{matrix} +0.38\\ -0.89\end{matrix}$} & {\textbf{2.10}$ \begin{matrix} +1.57\\ -0.73\end{matrix}$} \\
					\hline
				\end{tabular}
			\label{tab9}
		\end{center}
	\end{table}

	The features of malicious traffic in HMCT-2020(19-20) have no major change in the overall spatio-temporal behavior compared to HMCT-2020(18) and only the attack information in packets is different.
	N-gram+SVM relies more on the content information in packets and works well when detecting malicious traffic similar to the training data in contents.
	When the information changes, its generalization performance will decrease significantly.
	IG-PCA-Ensemble relies more on the behavior characteristics of malicious traffic at Flow-level and is not sensitive to attack information, which results in its overall detection performance being centered.
	Compared with these two methods, HMCD-Model extracts hierarchical spatio-temporal of traffic features from Pkt-level and Flow-level respectively. Therefore, it shows excellent detection performance to resist concept migration.

\subsubsection{Generalization Comparison in Dataset CIC-IDS-2017}


We selected the data in the dataset HMCT-2020(18) and CIC-IDS-2017 for generalization detection, one for training and the other for testing.
Since the malicious communication traffic type in CIC-IDS-2017 is relatively single and the correspondingly constructed field-based dictionary cannot cover more malicious scenes, making it difficult for HMCD-Model to generate GAFs.
Therefore, we use data in HMCT-2020(18) as the training set and data in CIC-IDS-2017 as the test set.
The experimental results are shown in Tab~\ref{tab8}.
The precision of N-gram+SVM is $ P= $98.38\% and the $ FPR $ is 0.35, but other metrics are very low. This is because N-gram+SVM increases the threshold for determining a sample as malicious, which means a lot of false negatives.
For instance, the $ F1 $ of N-gram+SVM is 56.34\%, and is 34.35\% lower than HMCD-Model. In addition, N-gram+SVM and IG-PCA-Ensemble have large fluctuations in CIC-IDS-2017, which makes them easy to lose detection performance. 
Therefore, in general, our method has the best overall performance. %

\begin{table*}[htbp]
	\caption{Generalization comparison in dataset CIC-IDS-2017 (ep3 in Tab~\ref{tab123})}
	\begin{center}
			\begin{tabular}{|c|c|c|c|c|}
				\hline
				\multirow{2}{*}{\diagbox[height=35pt,innerrightsep=40pt]{\makecell[l]{\ }}{\ }}& \multicolumn{4}{c|}{\tabincell{c}{CIC-IDS-2017 \\ (ep3)}} \\ 
				\cline{2-5} 
				& $ P(\%) $ & $ R(\%) $ & $ F1(\%) $ & $ FPR(\%) $ \\
				\hline
				N-gram+SVM & {\textbf{98.38}$ \begin{matrix} +1.31\\ -2.44\end{matrix}$} & {51.28$ \begin{matrix} +43.55\\ -48.92\end{matrix}$} & {56.34$ \begin{matrix} +40.86\\ -51.73\end{matrix}$} & {\textbf{0.35}$ \begin{matrix} +0.53\\ -0.30\end{matrix}$} \\
				\hline
				IG-PCA-Ensemble  & {73.71$ \begin{matrix} +15.68\\ -62.25\end{matrix}$} & {75.04$ \begin{matrix} +18.53\\ -74.06\end{matrix}$} & {73.45$ \begin{matrix} +17.98\\ -71.65\end{matrix}$} & {10.50$ \begin{matrix} +0.96\\ -2.94\end{matrix}$} \\
				\hline
				HMCD-Model & {86.93$ \begin{matrix} +12.03\\ -11.13\end{matrix}$} & {\textbf{95.32}$ \begin{matrix} +3.14\\ -2.80\end{matrix}$} & {\textbf{90.69}$ \begin{matrix} +6.77\\ -5.03\end{matrix}$} & {15.37$ \begin{matrix} +16.06\\ -14.37\end{matrix}$} \\
				\hline
			\end{tabular}
		\label{tab8}
	\end{center}
	\vspace{-1.5em}
\end{table*}

	The detection results of these three methods have large fluctuations.
	The fluctuation of N-gram+SVM comes from the change of training data.
	N-gram+SVM is sensitive to content information of packets.
	It is difficult to extract features effectively when the difference between the content information of the training data and the test data exceeds a threshold. 
	The fluctuation of IG-PCA-Ensemble mainly comes from the change of training data and the uncertainty of its own model.
	The fluctuation of HMCD-Model mainly comes from the initialization and update of parameters of the model itself. But it can be stabilized through multiple iterations of training, so the HMCD-Model is relatively more stable.

\subsubsection{Generalization Comparison in 82-Malware-Traffic}

	
	We capture relevant malicious traffic data from the Internet to verify the generalization performance of the model more comprehensively.
	We use data in HMCT-2020(18) as the training set and data in 82-Malware-Traffic as the test set.
	The experimental results are shown in Tab~\ref{tab7}.
	Similarly, the precision of N-gram+SVM is $ P= $99.71\% and the $ FPR $ is 0.14, but other metrics are very low. This is because N-gram+SVM increases the threshold for determining a sample as malicious, which means a lot of false negatives.
	For instance, the $ F1 $ of N-gram+SVM is 66.09\%, and is 17.57\% lower than HMCD-Model.
	In general, the HMCD-Model also has the best comprehensive performance and is more stable, with $ F1 $ of 83.66\% and $ FPR $ of 2.57\%.
	
	\begin{table}[htbp]
		\caption{Generalization comparison in 82-Malware-Traffic (ep4 in Tab~\ref{tab123})}
		\begin{center}
				\begin{tabular}{|c|c|c|c|c|}
					\hline
					\multirow{2}{*}{\diagbox[height=35pt,innerrightsep=40pt]{\makecell[l]{\ }}{\ }}& \multicolumn{4}{c|}{\tabincell{c}{82-Malware-Traffic \\ (ep4)}} \\ 
					\cline{2-5} 
					& $ P(\%)$ & $ R(\%) $ & $ F1(\%) $ & $ FPR(\%) $ \\
					\hline
					N-gram+SVM & {\textbf{99.71}$ \begin{matrix} +0.16\\ -0.26\end{matrix}$} & {49.43$ \begin{matrix} +0.29\\ -1.09\end{matrix}$} & {66.09$ \begin{matrix} +0.26\\ -0.94\end{matrix}$} & {\textbf{0.14}$ \begin{matrix} +0.14\\ -0.08\end{matrix}$} \\
					\hline
					IG-PCA-Ensemble & {94.18$ \begin{matrix} +0.45\\ -0.64\end{matrix}$} & {72.55$ \begin{matrix} +14.03\\ -20.92\end{matrix}$} & {80.81$ \begin{matrix} +9.15\\ -14.00\end{matrix}$} & {4.54$ \begin{matrix} +1.44\\ -1.61\end{matrix}$} \\
					\hline
					HMCD-Model & {96.62$ \begin{matrix} +2.26\\ -4.27\end{matrix}$} & {\textbf{73.77}$ \begin{matrix} +2.01\\ -3.41\end{matrix}$} & {\textbf{83.66}$ \begin{matrix} +2.15\\ -3.79\end{matrix}$} & {2.57$ \begin{matrix} +3.26\\ -1.71\end{matrix}$} \\
					\hline
				\end{tabular}
			\label{tab7}
		\end{center}
	\end{table}
	
	The detection results of the three detection models in the real traffic environment has deteriorated compared with in HMCT-2020 and CIC-IDS-2017.
	This is because the real network environment is more complicated and the form of unknown malicious traffic is more variable.
	The traffic generated by malware is not all traffic related to malicious communication behavior, which leads to false positives.

\subsubsection{Comparison in Time Cost}

\begin{figure}[ht]
	\centerline{\includegraphics[scale=0.6]{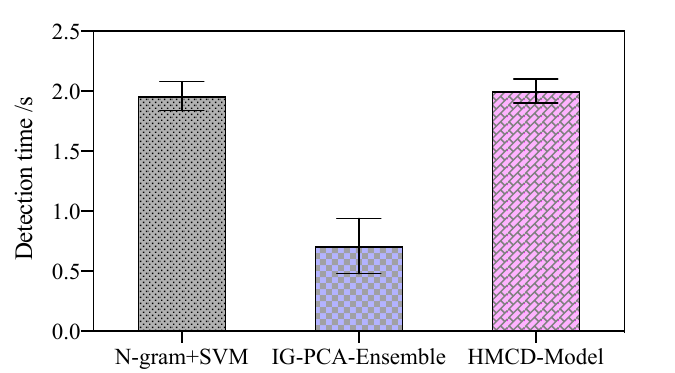}}
	\caption{Comparison on detection time cost (per 10,000 samples).}
	\label{time-fig}
\end{figure}

We compared the detection time of HMCD-Model with N-gram+SVM and IG-PCA-Ensemble. Taking 10,000 samples as a unit, the detection time of each model is shown in Fig~\ref{time-fig}. IG-PCA-Ensemble spends the least time in detecting because its feature space is low-dimensional. However, there is no free lunch in the world. Although the detection time of IG-PCA-Ensemble is fast, its generalization performance is not satisfactory.

In addition, N-gram+SVM spends the second least time in detecting, because its feature space is relatively small and it uses linear SVM for detection.
However, N-gram+SVM uses bag-of-words and Chi-square algorithms in pre-processing, which means it is difficult to arbitrarily increase the packet extraction content and feature space.
Its storage space, pre-processing time and SVM convergence time will increase exponentially when there is too much input information. 

The detection time of HMCD-Model is of the same magnitude as N-gram+SVM.
In fact, the internal structure of HMCD-Model supports parallel processing, such as the processing of packet text and statistical features can be performed in parallel.
The matrix operation characteristics of neural networks also enable HMCD-Model to make full use of the computing advantages of GPUs.
More importantly, HMCD-Model maintains a good generalization detection ability, the detection time and the size of the feature space of the data are linear.
Therefore, HCMD-Model is undoubtedly a more competitive and better comprehensive performance detection method from the current Internet big data environment. 

\section{Discussion}


\textbf{1) How to select the hyper-parameters?}

We determine some of the best hyper-parameters (the number of packets in a flow, the byte size in a packet, \textit{etc.}) of HMCT-Model through experiments.
The magnitude of the combination of hyper-parameters is very high and we cannot evaluate all solutions.
As a rule of thumb, we use the method of controlling variables to cover the best solution as much as possible and to achieve a trade-off between detection performance and time cost.
For instance, increasing the number of packets in a flow will improve the detection accuracy of the model, but the convergence time and detection speed will be slower.
Therefore, we set two packets in a flow considering the trade-off between detection time and accuracy. 

\textbf{2) What are the limitations of HMCD-Model?}

HMCD-Model will cause misjudgment in two cases:
First, some packets (such as packets in CIC-IDS-2017) have less packet content and insufficient features. Fewer features cannot give the model enough information.
Second, the traffic generated by malware is indistinguishable from benign data.
For instance, some malicious request packets have only two lines.
However, benign packets in practice also have similar characteristics, which leads to inaccurate final judgment results of the model. 

GAN can be used to generate HTTP-based malicious communication traffic, but it is generated based on existing frameworks.
It still has certain limitations in many unknown attack scenes.

	
\textbf{3) Why choose WGAN-GP to generate GAFs?}

	Compared with vanilla GAN, WGAN-GP has stronger imitation ability and easier convergence, while maintaining the diversity of generated data \cite{gulrajani2017improved}. HTTP-based application layer traffic belongs to text data, which is more difficult to generate than image data, and vanilla GAN is difficult to imitate. In this paper, when we try to generate GAFs with vanilla GAN, the model has a high probability of collapse and fails to converge. Therefore, WGAN-GP is relatively more suitable.
	

\textbf{4) Can GAFs be used in other related methods to improve performance?} 

	In this paper, we utilize WGAN-GP to construct generated adversarial flows (GAFs) for data enhancement, to compensate for the insufficiency of real HTTP-based malicious communication traffic.
	The experiments in Section 6.3.2 show that GAFs can improve the generalization of HMCD-Model. We also try to use GAFs for other related methods.
	However, experimental results show that although GAFs can improve the performance of other methods, the improvement is not much, and sometimes even leads to performance degradation.
	There are two reasons here. First, other related methods have limitations in detecting HTTP-based malicious communication traffic and are difficult to comprehensively extract relevant features.
	Second, GAFs have a fixed composition, a flow consists of only one request and one response packet, and the represented feature space is limited.
	
	Taking HMCT2020 (19-20) as an example, we use ep2 as the training set and test set, and increase GAFs when training N-gram+SVM and IG-PCA-Ensemble. The experimental results are shown in Tab~\ref{tab9x}. For N-gram+SVM, after adding GAFs in the training set, the precision rate is greatly improved (81.79\% to 87.56\%), but the recall rate is reduced (62.99\% to 63.33\%). For IG-PCA-Ensemble, after adding GAFs in the training set, the precision rate decreases (75.92\% to 73.85\%) and the recall rate increases (70.07\% to 78.55\%). In general, the improvement effect of GAFs on other methods is limited, and for HMCD-Model, the generalization can be improved comprehensively.

	\begin{table}[htbp]
	\caption{Generalization Comparison with GAFs in dataset HMCT-2020(19-20) (ep2 in Tab~\ref{tab123})} 
	\begin{center}
			\begin{tabular}{|c|c|c|c|c|}
				\hline
				\multirow{2}{*}{\diagbox[height=35pt,innerrightsep=49pt]{\makecell[l]{\ \ }}{\ \ }} & \multicolumn{4}{c|}{\tabincell{c}{HMCT-2020(19-20) \\ (ep2)}} \\
				\cline{2-5} 
				& $ P(\%) $ & $ R(\%) $ & $ F1(\%) $ & $ FPR(\%) $ \\
				\hline
				N-gram+SVM & {81.79$ \begin{matrix} +1.83\\ -1.07\end{matrix}$} & {51.24$ \begin{matrix} +1.31\\ -1.24\end{matrix}$} & {62.99$ \begin{matrix} +0.73\\ -0.84\end{matrix}$} & {11.42$ \begin{matrix} +1.08\\ -1.34\end{matrix}$} \\
				\hline
				N-gram+SVM (\textbf{+GAFs})& {87.56$ \begin{matrix} +0.81\\ -2.16\end{matrix}$} & {49.60$ \begin{matrix} +2.09\\ -1.82\end{matrix}$} & {63.33$ \begin{matrix} +0.73\\ -0.84\end{matrix}$} & {6.75$ \begin{matrix} +1.04\\ -0.98\end{matrix}$} \\
				\hline
				IG-PCA-Ensemble & {75.92$ \begin{matrix} +6.05\\ -6.21\end{matrix}$} & {70.07$ \begin{matrix} +4.68\\ -4.87\end{matrix}$} & {72.59$ \begin{matrix} +0.60\\ -0.45\end{matrix}$} & {23.02$ \begin{matrix} +9.46\\ -8.68\end{matrix}$} \\
				\hline
				IG-PCA-Ensemble (\textbf{+GAFs}) & {73.85$ \begin{matrix} +5.45\\ -6.52\end{matrix}$} & {78.55$ \begin{matrix} +3.81\\ -4.93\end{matrix}$} & {76.13$ \begin{matrix} +0.74\\ -0.58\end{matrix}$} & {24.96$ \begin{matrix} +8.52\\ -9.79\end{matrix}$} \\
				\hline
				HMCD-Model & {\textbf{97.94}$ \begin{matrix} +0.70\\ -1.51\end{matrix}$} & {\textbf{99.39}$ \begin{matrix} +0.18\\ -0.25\end{matrix}$} & \textbf{{98.66}$ \begin{matrix} +0.38\\ -0.89\end{matrix}$} & {\textbf{2.10}$ \begin{matrix} +1.57\\ -0.73\end{matrix}$} \\
				\hline
			\end{tabular}
		\label{tab9x}
	\end{center}
\end{table}

\textbf{5) Can HMCD-Model be used for HTTPS encrypted traffic detection?}

Currently, many studies have achieved good results in protocol classification and application identification \cite{shen2017classification,di2019realistically,lotfollahi2020deep} based on the strategy that directly analyzing encrypted traffic (such as HTTPS) without decryption.
However, these methods are difficult to be effective for the refined detection of highly complex and covert attack traffic (such as HTTP-based malicious communication traffic).
The current mainstream strategy is to deploy the detection system in the terminal instead of the middle.
In this way, the natural decryption capability of the terminal can be used to convert HTTPS to HTTP.
Many methods (Wang \emph{et al.} \cite{wang2017detecting}, \textit{etc.}) and systems (WAF \cite{clincy2018web}, \textit{etc.}) for malicious behavior detection are based on this strategy.
We can decrypt HTTPS traffic into HTTP traffic and analyze the malicious behavior it contains.
HMCD-Model based on the above strategy can be feasible as an end-to-end detection method. It can be applied to the IDSs of host/server. 

\section{Conclusion}
For the detection of unknown HTTP-based malicious communication behavior, we propose HMCD-Model. We build a hybrid neural network to learn the hierarchical spatial-temporal features of traffic; use GAN to generate adversarial flows to make up for the lack of real traffic.
In addition, we publish dataset HMCT-2020. Compared with the most representative methods at present, our model has obvious advantages in generalization, which can reach the $ F1 \approx 83.66\% $ in detection of real malware traffic. HMCD-Model can be applied to the detection of unknown HTTP-based malicious communication behavior and improve the capability of defenders against highly complex and covert network attack events. 

In the future, we will further collect and expand datasets. In addition, we will fine-grain such malicious behavior scene to improve the ability of GAN-based malicious traffic generation.

\section*{Acknowledgements}
	This work is supported by the National Key Research and Development Program of China \\ (Grant No.2018YFB0804704), and the National Key Research and Development Program of \\ China (Grant No.2019YFB1005201).

\appendices

\section{Generalization Comparison with Classical Baseline Methods}


	We compared HMCD-Model with classical machine learning methods (Bayes, Decision Tree, SVM) and conventional neural networks (DNN, CNN, LSTM).
	We use data in HMCT-2020(18) as the training set and data in 82-Malware-Traffic as the test set.	The experimental results are shown in Tab.~\ref{tab6}.
	HMCD-Model is at least 5.53\% better than other methods in terms of $ F1 $.
	Although Naive Bayes can achieve the $ R $ of 82.07\%, the $ FPR $ is the highest, 28.03\%. This is because Naive Bayes decreases the threshold for determining a sample as malicious, which means a lot of false positives.
	Similarly, the precision of SVM is $ P= $99.87\% and the $ FPR $ is 0.08, but other metrics are very low. This is because SVM increases the threshold for determining a sample as malicious, which means a lot of false negatives. For instance, the $ F1 $ of SVM is only 75.16\%.
	Therefore, HMCD-Model has a better comprehensive detection performance than other classical methods.

	\begin{table}[htbp]
	\caption{Comparison with classical baseline methods (ep4 in Tab~\ref{tab123})}
	\begin{center}
			\begin{tabular}{|c|c|c|c|c|}
				\hline
				\multirow{2}{*}{\diagbox[height=35pt,innerrightsep=40pt]{\makecell[l]{\ }}{\ }}& \multicolumn{4}{c|}{\tabincell{c}{82-Malware-Traffic \\ (ep4)}} \\ 
				\cline{2-5} 
				& $ P(\%)$ & $ R(\%) $ & $ F1(\%) $ & $ FPR(\%) $ \\
				\hline
				Naive Bayes & {74.55$ \begin{matrix} +0.52\\ -0.26\end{matrix}$} & {\textbf{82.07}$ \begin{matrix} +0.08\\ -0.2\end{matrix}$} & {78.13$ \begin{matrix} +0.32\\ -0.12\end{matrix}$} & {28.03$ \begin{matrix} +0.4\\ -0.75\end{matrix}$} \\
				\hline
				\tabincell{c}{Decision Tree} & {99.7$ \begin{matrix} +0.13\\ -0.23\end{matrix}$} & {58.08$ \begin{matrix} +1.77\\ -4.06\end{matrix}$} & {73.37$ \begin{matrix} +1.46\\ -3.28\end{matrix}$} & {0.18$ \begin{matrix} +0.14\\ -0.08\end{matrix}$} \\
				\hline
				SVM & {\textbf{99.87}$ \begin{matrix} +0.08\\ -0.08\end{matrix}$} & {60.26$ \begin{matrix} +0.77\\ -0.48\end{matrix}$} & {75.16$ \begin{matrix} +0.57\\ -0.35\end{matrix}$} & {\textbf{0.08}$ \begin{matrix} +0.05\\ -0.05\end{matrix}$} \\
				\hline
				DNN & {99.83$ \begin{matrix} +0.06\\ -0.06\end{matrix}$} & {55.82$ \begin{matrix} +0.59\\ -0.5\end{matrix}$} & {71.6$ \begin{matrix} +0.49\\ -0.41\end{matrix}$} & {0.10$ \begin{matrix} +0.03\\ -0.04\end{matrix}$} \\
				\hline
				CNN & {94.95$ \begin{matrix} +1.92\\ -4.45\end{matrix}$} & {52.0$ \begin{matrix} +2.37\\ -2.16\end{matrix}$} & {67.15$ \begin{matrix} +1.38\\ -1.51\end{matrix}$} & {2.83$ \begin{matrix} +2.87\\ -1.14\end{matrix}$} \\
				\hline
				LSTM & {99.77$ \begin{matrix} +0.11\\ -0.06\end{matrix}$} & {55.44$ \begin{matrix} +2.62\\ -1.27\end{matrix}$} & {71.26$ \begin{matrix} +2.13\\ -1.03\end{matrix}$} & {0.13$ \begin{matrix} +0.03\\ -0.07\end{matrix}$} \\
				\hline
				\tabincell{c}{HMCD-Model} & {96.62$ \begin{matrix} +2.26\\ -4.27\end{matrix}$} & {73.77$ \begin{matrix} +2.01\\ -3.41\end{matrix}$} & {\textbf{83.66}$ \begin{matrix} +2.15\\ -3.79\end{matrix}$} & {2.57$ \begin{matrix} +3.26\\ -1.71\end{matrix}$} \\
				\hline
		\end{tabular}
		\label{tab6}
	\end{center}
	\end{table}

\bibliographystyle{cas-model2-names}

\bibliography{sample}

\begin{thebibliography}{49}
\expandafter\ifx\csname natexlab\endcsname\relax\def\natexlab#1{#1}\fi
\providecommand{\url}[1]{\texttt{#1}}
\providecommand{\href}[2]{#2}
\providecommand{\path}[1]{#1}
\providecommand{\DOIprefix}{doi:}
\providecommand{\ArXivprefix}{arXiv:}
\providecommand{\URLprefix}{URL: }
\providecommand{\Pubmedprefix}{pmid:}
\providecommand{\doi}[1]{\href{http://dx.doi.org/#1}{\path{#1}}}
\providecommand{\Pubmed}[1]{\href{pmid:#1}{\path{#1}}}
\providecommand{\bibinfo}[2]{#2}
\ifx\xfnm\relax \def\xfnm[#1]{\unskip,\space#1}\fi
\bibitem[{Aburomman and Reaz(2017)}]{aburomman2017survey}
\bibinfo{author}{Aburomman, A.A.}, \bibinfo{author}{Reaz, M.B.I.},
  \bibinfo{year}{2017}.
\newblock \bibinfo{title}{A survey of intrusion detection systems based on
  ensemble and hybrid classifiers}.
\newblock \bibinfo{journal}{Computers \& Security} \bibinfo{volume}{65},
  \bibinfo{pages}{135--152}.
\bibitem[{Altman(1992)}]{altman1992introduction}
\bibinfo{author}{Altman, N.S.}, \bibinfo{year}{1992}.
\newblock \bibinfo{title}{An introduction to kernel and nearest-neighbor
  nonparametric regression}.
\newblock \bibinfo{journal}{The American Statistician} \bibinfo{volume}{46},
  \bibinfo{pages}{175--185}.
\bibitem[{Cannady(1998)}]{cannady1998artificial}
\bibinfo{author}{Cannady, J.}, \bibinfo{year}{1998}.
\newblock \bibinfo{title}{Artificial neural networks for misuse detection}, in:
  \bibinfo{booktitle}{National information systems security conference},
  \bibinfo{organization}{Baltimore}. pp. \bibinfo{pages}{443--456}.
\bibitem[{Caviglione et~al.(2015)Caviglione, Gaggero, Lalande, Mazurczyk and
  Urba{\'n}ski}]{caviglione2015seeing}
\bibinfo{author}{Caviglione, L.}, \bibinfo{author}{Gaggero, M.},
  \bibinfo{author}{Lalande, J.F.}, \bibinfo{author}{Mazurczyk, W.},
  \bibinfo{author}{Urba{\'n}ski, M.}, \bibinfo{year}{2015}.
\newblock \bibinfo{title}{Seeing the unseen: revealing mobile malware hidden
  communications via energy consumption and artificial intelligence}.
\newblock \bibinfo{journal}{IEEE Transactions on Information Forensics and
  Security} \bibinfo{volume}{11}, \bibinfo{pages}{799--810}.
\bibitem[{Chandola et~al.(2009)Chandola, Banerjee and
  Kumar}]{chandola2009anomaly}
\bibinfo{author}{Chandola, V.}, \bibinfo{author}{Banerjee, A.},
  \bibinfo{author}{Kumar, V.}, \bibinfo{year}{2009}.
\newblock \bibinfo{title}{Anomaly detection: A survey}.
\newblock \bibinfo{journal}{ACM computing surveys (CSUR)} \bibinfo{volume}{41},
  \bibinfo{pages}{1--58}.
\bibitem[{Chen and Bates(1996)}]{chen1996rfc1998}
\bibinfo{author}{Chen, E.}, \bibinfo{author}{Bates, T.}, \bibinfo{year}{1996}.
\newblock \bibinfo{title}{Rfc1998: An application of the bgp community
  attribute in multi-home routing}.
\bibitem[{Cheng(2019)}]{cheng2019pac}
\bibinfo{author}{Cheng, A.}, \bibinfo{year}{2019}.
\newblock \bibinfo{title}{Pac-gan: Packet generation of network traffic using
  generative adversarial networks}, in: \bibinfo{booktitle}{2019 IEEE 10th
  Annual Information Technology, Electronics and Mobile Communication
  Conference (IEMCON)}, \bibinfo{organization}{IEEE}. pp.
  \bibinfo{pages}{0728--0734}.
\bibitem[{Cheng et~al.(2021)Cheng, Zhou, Shen, Kong and Wu}]{cheng2021packet}
\bibinfo{author}{Cheng, Q.}, \bibinfo{author}{Zhou, S.}, \bibinfo{author}{Shen,
  Y.}, \bibinfo{author}{Kong, D.}, \bibinfo{author}{Wu, C.},
  \bibinfo{year}{2021}.
\newblock \bibinfo{title}{Packet-level adversarial network traffic crafting
  using sequence generative adversarial networks}.
\newblock \bibinfo{journal}{arXiv preprint arXiv:2103.04794} .
\bibitem[{Chowdhury et~al.(2017)Chowdhury, Hammond, Konowicz, Xin, Wu and
  Li}]{chowdhury2017few}
\bibinfo{author}{Chowdhury, M.M.U.}, \bibinfo{author}{Hammond, F.},
  \bibinfo{author}{Konowicz, G.}, \bibinfo{author}{Xin, C.},
  \bibinfo{author}{Wu, H.}, \bibinfo{author}{Li, J.}, \bibinfo{year}{2017}.
\newblock \bibinfo{title}{A few-shot deep learning approach for improved
  intrusion detection}, in: \bibinfo{booktitle}{2017 IEEE 8th Annual Ubiquitous
  Computing, Electronics and Mobile Communication Conference (UEMCON)},
  \bibinfo{organization}{IEEE}. pp. \bibinfo{pages}{456--462}.
\bibitem[{Clincy and Shahriar(2018)}]{clincy2018web}
\bibinfo{author}{Clincy, V.}, \bibinfo{author}{Shahriar, H.},
  \bibinfo{year}{2018}.
\newblock \bibinfo{title}{Web application firewall: Network security models and
  configuration}, in: \bibinfo{booktitle}{2018 IEEE 42nd Annual Computer
  Software and Applications Conference (COMPSAC)},
  \bibinfo{organization}{IEEE}. pp. \bibinfo{pages}{835--836}.
\bibitem[{Di~Martino et~al.(2019)Di~Martino, Quax and
  Lamotte}]{di2019realistically}
\bibinfo{author}{Di~Martino, M.}, \bibinfo{author}{Quax, P.},
  \bibinfo{author}{Lamotte, W.}, \bibinfo{year}{2019}.
\newblock \bibinfo{title}{Realistically fingerprinting social media webpages in
  https traffic}, in: \bibinfo{booktitle}{Proceedings of the 14th International
  Conference on Availability, Reliability and Security}, pp.
  \bibinfo{pages}{1--10}.
\bibitem[{Du et~al.(2017)Du, Li, Zheng and Srikumar}]{du2017deeplog}
\bibinfo{author}{Du, M.}, \bibinfo{author}{Li, F.}, \bibinfo{author}{Zheng,
  G.}, \bibinfo{author}{Srikumar, V.}, \bibinfo{year}{2017}.
\newblock \bibinfo{title}{Deeplog: Anomaly detection and diagnosis from system
  logs through deep learning}, in: \bibinfo{booktitle}{Proceedings of the 2017
  ACM SIGSAC Conference on Computer and Communications Security}, pp.
  \bibinfo{pages}{1285--1298}.
\bibitem[{Du et~al.(2018)Du, Ma, Li, Li, Sun and Liu}]{du2018network}
\bibinfo{author}{Du, Z.}, \bibinfo{author}{Ma, L.}, \bibinfo{author}{Li, H.},
  \bibinfo{author}{Li, Q.}, \bibinfo{author}{Sun, G.}, \bibinfo{author}{Liu,
  Z.}, \bibinfo{year}{2018}.
\newblock \bibinfo{title}{Network traffic anomaly detection based on wavelet
  analysis}, in: \bibinfo{booktitle}{2018 IEEE 16th International Conference on
  Software Engineering Research, Management and Applications (SERA)},
  \bibinfo{organization}{IEEE}. pp. \bibinfo{pages}{94--101}.
\bibitem[{Fukushima et~al.(1983)Fukushima, Miyake and Ito}]{Neocognitron}
\bibinfo{author}{Fukushima, K.}, \bibinfo{author}{Miyake, S.},
  \bibinfo{author}{Ito, T.}, \bibinfo{year}{1983}.
\newblock \bibinfo{title}{Neocognitron: A neural network model for a mechanism
  of visual pattern recognition}.
\newblock \bibinfo{journal}{IEEE transactions on systems, man, and cybernetics}
  , \bibinfo{pages}{826--834}.
\bibitem[{Ghafir et~al.(2015)Ghafir, Svoboda, Prenosil
  et~al.}]{ghafir2015survey}
\bibinfo{author}{Ghafir, I.}, \bibinfo{author}{Svoboda, J.},
  \bibinfo{author}{Prenosil, V.}, et~al., \bibinfo{year}{2015}.
\newblock \bibinfo{title}{A survey on botnet command and control traffic
  detection}.
\newblock \bibinfo{journal}{Int J Adv Comput Netw Secur} \bibinfo{volume}{5},
  \bibinfo{pages}{7580}.
\bibitem[{Gulrajani et~al.(2017)Gulrajani, Ahmed, Arjovsky, Dumoulin and
  Courville}]{gulrajani2017improved}
\bibinfo{author}{Gulrajani, I.}, \bibinfo{author}{Ahmed, F.},
  \bibinfo{author}{Arjovsky, M.}, \bibinfo{author}{Dumoulin, V.},
  \bibinfo{author}{Courville, A.C.}, \bibinfo{year}{2017}.
\newblock \bibinfo{title}{Improved training of wasserstein gans}, in:
  \bibinfo{booktitle}{Advances in neural information processing systems}, pp.
  \bibinfo{pages}{5767--5777}.
\bibitem[{Gupta et~al.(2019)Gupta, Khanna, SK, Shankar, Furtado and
  Rodrigues}]{gupta2019efficient}
\bibinfo{author}{Gupta, D.}, \bibinfo{author}{Khanna, A.}, \bibinfo{author}{SK,
  L.}, \bibinfo{author}{Shankar, K.}, \bibinfo{author}{Furtado, V.},
  \bibinfo{author}{Rodrigues, J.J.}, \bibinfo{year}{2019}.
\newblock \bibinfo{title}{Efficient artificial fish swarm based clustering
  approach on mobility aware energy-efficient for manet}.
\newblock \bibinfo{journal}{Transactions on Emerging Telecommunications
  Technologies} \bibinfo{volume}{30}, \bibinfo{pages}{e3524}.
\bibitem[{Hajisalem and Babaie(2018)}]{hajisalem2018hybrid}
\bibinfo{author}{Hajisalem, V.}, \bibinfo{author}{Babaie, S.},
  \bibinfo{year}{2018}.
\newblock \bibinfo{title}{A hybrid intrusion detection system based on abc-afs
  algorithm for misuse and anomaly detection}.
\newblock \bibinfo{journal}{Computer Networks} \bibinfo{volume}{136},
  \bibinfo{pages}{37--50}.
\bibitem[{Hao et~al.(2021)Hao, Jiang, Xiao, Wang, Yao, Liu and
  Liu}]{hao2021producing}
\bibinfo{author}{Hao, X.}, \bibinfo{author}{Jiang, Z.}, \bibinfo{author}{Xiao,
  Q.}, \bibinfo{author}{Wang, Q.}, \bibinfo{author}{Yao, Y.},
  \bibinfo{author}{Liu, B.}, \bibinfo{author}{Liu, J.}, \bibinfo{year}{2021}.
\newblock \bibinfo{title}{Producing more with less: A gan-based network attack
  detection approach for imbalanced data}, in: \bibinfo{booktitle}{2021 IEEE
  24th International Conference on Computer Supported Cooperative Work in
  Design (CSCWD)}, \bibinfo{organization}{IEEE}. pp. \bibinfo{pages}{384--390}.
\bibitem[{Hochreiter and Schmidhuber(1997)}]{Hochreiter}
\bibinfo{author}{Hochreiter, S.}, \bibinfo{author}{Schmidhuber, J.},
  \bibinfo{year}{1997}.
\newblock \bibinfo{title}{Long short-term memory}.
\newblock \bibinfo{journal}{Neural Computation} \bibinfo{volume}{9},
  \bibinfo{pages}{1735--1780}.
\bibitem[{Jan et~al.(2020)Jan, Hao, Hu, Pu, Oswal, Wang and
  Viswanath}]{jan2020throwing}
\bibinfo{author}{Jan, S.T.}, \bibinfo{author}{Hao, Q.}, \bibinfo{author}{Hu,
  T.}, \bibinfo{author}{Pu, J.}, \bibinfo{author}{Oswal, S.},
  \bibinfo{author}{Wang, G.}, \bibinfo{author}{Viswanath, B.},
  \bibinfo{year}{2020}.
\newblock \bibinfo{title}{Throwing darts in the dark? detecting bots with
  limited data using neural data augmentation}, in: \bibinfo{booktitle}{The
  41st IEEE Symposium on Security and Privacy (IEEE SP)}.
\bibitem[{Jose et~al.(2018)Jose, Malathi, Reddy and Jayaseeli}]{jose2018survey}
\bibinfo{author}{Jose, S.}, \bibinfo{author}{Malathi, D.},
  \bibinfo{author}{Reddy, B.}, \bibinfo{author}{Jayaseeli, D.},
  \bibinfo{year}{2018}.
\newblock \bibinfo{title}{A survey on anomaly based host intrusion detection
  system}, in: \bibinfo{booktitle}{Journal of Physics: Conference Series},
  \bibinfo{organization}{IOP Publishing}. p. \bibinfo{pages}{012049}.
\bibitem[{Karaboga and Basturk(2008)}]{karaboga2008performance}
\bibinfo{author}{Karaboga, D.}, \bibinfo{author}{Basturk, B.},
  \bibinfo{year}{2008}.
\newblock \bibinfo{title}{On the performance of artificial bee colony (abc)
  algorithm}.
\newblock \bibinfo{journal}{Applied soft computing} \bibinfo{volume}{8},
  \bibinfo{pages}{687--697}.
\bibitem[{Kim et~al.(2018)Kim, Kang, Rho, Sezer and Im}]{kim2018multimodal}
\bibinfo{author}{Kim, T.}, \bibinfo{author}{Kang, B.}, \bibinfo{author}{Rho,
  M.}, \bibinfo{author}{Sezer, S.}, \bibinfo{author}{Im, E.G.},
  \bibinfo{year}{2018}.
\newblock \bibinfo{title}{A multimodal deep learning method for android malware
  detection using various features}.
\newblock \bibinfo{journal}{IEEE Transactions on Information Forensics and
  Security} \bibinfo{volume}{14}, \bibinfo{pages}{773--788}.
\bibitem[{Kwon et~al.(2019)Kwon, Kim, Kim, Suh, Kim and Kim}]{kwon2019survey}
\bibinfo{author}{Kwon, D.}, \bibinfo{author}{Kim, H.}, \bibinfo{author}{Kim,
  J.}, \bibinfo{author}{Suh, S.C.}, \bibinfo{author}{Kim, I.},
  \bibinfo{author}{Kim, K.J.}, \bibinfo{year}{2019}.
\newblock \bibinfo{title}{A survey of deep learning-based network anomaly
  detection}.
\newblock \bibinfo{journal}{Cluster Computing} , \bibinfo{pages}{1--13}.
\bibitem[{Li et~al.(2019)Li, Zhou, Li, Yan and Zhu}]{li2019dynamic}
\bibinfo{author}{Li, J.}, \bibinfo{author}{Zhou, L.}, \bibinfo{author}{Li, H.},
  \bibinfo{author}{Yan, L.}, \bibinfo{author}{Zhu, H.}, \bibinfo{year}{2019}.
\newblock \bibinfo{title}{Dynamic traffic feature camouflaging via generative
  adversarial networks}, in: \bibinfo{booktitle}{2019 IEEE Conference on
  Communications and Network Security (CNS)}, \bibinfo{organization}{IEEE}. pp.
  \bibinfo{pages}{268--276}.
\bibitem[{Lin et~al.(2018)Lin, Shi and Xue}]{lin2018idsgan}
\bibinfo{author}{Lin, Z.}, \bibinfo{author}{Shi, Y.}, \bibinfo{author}{Xue,
  Z.}, \bibinfo{year}{2018}.
\newblock \bibinfo{title}{Idsgan: Generative adversarial networks for attack
  generation against intrusion detection}.
\newblock \bibinfo{journal}{arXiv preprint arXiv:1809.02077} .
\bibitem[{Lotfollahi et~al.(2020)Lotfollahi, Siavoshani, Zade and
  Saberian}]{lotfollahi2020deep}
\bibinfo{author}{Lotfollahi, M.}, \bibinfo{author}{Siavoshani, M.J.},
  \bibinfo{author}{Zade, R.S.H.}, \bibinfo{author}{Saberian, M.},
  \bibinfo{year}{2020}.
\newblock \bibinfo{title}{Deep packet: A novel approach for encrypted traffic
  classification using deep learning}.
\newblock \bibinfo{journal}{Soft Computing} \bibinfo{volume}{24},
  \bibinfo{pages}{1999--2012}.
\bibitem[{Maki and Feller(1989)}]{maki1989intrusion}
\bibinfo{author}{Maki, M.C.}, \bibinfo{author}{Feller, W.J.},
  \bibinfo{year}{1989}.
\newblock \bibinfo{title}{Intrusion detection system}.
\newblock \bibinfo{note}{US Patent 4,879,544}.
\bibitem[{Moustafa and Slay(2015)}]{moustafa2015unsw}
\bibinfo{author}{Moustafa, N.}, \bibinfo{author}{Slay, J.},
  \bibinfo{year}{2015}.
\newblock \bibinfo{title}{Unsw-nb15: a comprehensive data set for network
  intrusion detection systems (unsw-nb15 network data set)}, in:
  \bibinfo{booktitle}{2015 military communications and information systems
  conference (MilCIS)}, \bibinfo{organization}{IEEE}. pp.
  \bibinfo{pages}{1--6}.
\bibitem[{Ring et~al.(2019)Ring, Schl{\"o}r, Landes and Hotho}]{ring2019flow}
\bibinfo{author}{Ring, M.}, \bibinfo{author}{Schl{\"o}r, D.},
  \bibinfo{author}{Landes, D.}, \bibinfo{author}{Hotho, A.},
  \bibinfo{year}{2019}.
\newblock \bibinfo{title}{Flow-based network traffic generation using
  generative adversarial networks}.
\newblock \bibinfo{journal}{Computers \& Security} \bibinfo{volume}{82},
  \bibinfo{pages}{156--172}.
\bibitem[{Ring et~al.(2017)Ring, Wunderlich, Gr{\"u}dl, Landes and
  Hotho}]{ring2017flow}
\bibinfo{author}{Ring, M.}, \bibinfo{author}{Wunderlich, S.},
  \bibinfo{author}{Gr{\"u}dl, D.}, \bibinfo{author}{Landes, D.},
  \bibinfo{author}{Hotho, A.}, \bibinfo{year}{2017}.
\newblock \bibinfo{title}{Flow-based benchmark data sets for intrusion
  detection}, in: \bibinfo{booktitle}{Proceedings of the 16th European
  conference on cyber warfare and security}, pp. \bibinfo{pages}{361--369}.
\bibitem[{Salo et~al.(2019)Salo, Nassif and Essex}]{salo2019dimensionality}
\bibinfo{author}{Salo, F.}, \bibinfo{author}{Nassif, A.B.},
  \bibinfo{author}{Essex, A.}, \bibinfo{year}{2019}.
\newblock \bibinfo{title}{Dimensionality reduction with ig-pca and ensemble
  classifier for network intrusion detection}.
\newblock \bibinfo{journal}{Computer Networks} \bibinfo{volume}{148},
  \bibinfo{pages}{164--175}.
\bibitem[{Selvakumar and Muneeswaran(2019)}]{selvakumar2019firefly}
\bibinfo{author}{Selvakumar, B.}, \bibinfo{author}{Muneeswaran, K.},
  \bibinfo{year}{2019}.
\newblock \bibinfo{title}{Firefly algorithm based feature selection for network
  intrusion detection}.
\newblock \bibinfo{journal}{Computers \& Security} \bibinfo{volume}{81},
  \bibinfo{pages}{148--155}.
\bibitem[{Sharafaldin et~al.(2018)Sharafaldin, Lashkari and
  Ghorbani}]{sharafaldin2018toward}
\bibinfo{author}{Sharafaldin, I.}, \bibinfo{author}{Lashkari, A.H.},
  \bibinfo{author}{Ghorbani, A.A.}, \bibinfo{year}{2018}.
\newblock \bibinfo{title}{Toward generating a new intrusion detection dataset
  and intrusion traffic characterization.}, in: \bibinfo{booktitle}{ICISSP},
  pp. \bibinfo{pages}{108--116}.
\bibitem[{Shen et~al.(2017)Shen, Wei, Zhu and Wang}]{shen2017classification}
\bibinfo{author}{Shen, M.}, \bibinfo{author}{Wei, M.}, \bibinfo{author}{Zhu,
  L.}, \bibinfo{author}{Wang, M.}, \bibinfo{year}{2017}.
\newblock \bibinfo{title}{Classification of encrypted traffic with second-order
  markov chains and application attribute bigrams}.
\newblock \bibinfo{journal}{IEEE Transactions on Information Forensics and
  Security} \bibinfo{volume}{12}, \bibinfo{pages}{1830--1843}.
\bibitem[{Shiravi et~al.(2012)Shiravi, Shiravi, Tavallaee and
  Ghorbani}]{shiravi2012toward}
\bibinfo{author}{Shiravi, A.}, \bibinfo{author}{Shiravi, H.},
  \bibinfo{author}{Tavallaee, M.}, \bibinfo{author}{Ghorbani, A.A.},
  \bibinfo{year}{2012}.
\newblock \bibinfo{title}{Toward developing a systematic approach to generate
  benchmark datasets for intrusion detection}.
\newblock \bibinfo{journal}{computers \& security} \bibinfo{volume}{31},
  \bibinfo{pages}{357--374}.
\bibitem[{Shrestha et~al.(2015)Shrestha, Hempel, Rezaei and
  Sharif}]{shrestha2015support}
\bibinfo{author}{Shrestha, P.L.}, \bibinfo{author}{Hempel, M.},
  \bibinfo{author}{Rezaei, F.}, \bibinfo{author}{Sharif, H.},
  \bibinfo{year}{2015}.
\newblock \bibinfo{title}{A support vector machine-based framework for
  detection of covert timing channels}.
\newblock \bibinfo{journal}{IEEE Transactions on Dependable and Secure
  Computing} \bibinfo{volume}{13}, \bibinfo{pages}{274--283}.
\bibitem[{Song et~al.(2011)Song, Takakura, Okabe, Eto, Inoue and
  Nakao}]{song2011statistical}
\bibinfo{author}{Song, J.}, \bibinfo{author}{Takakura, H.},
  \bibinfo{author}{Okabe, Y.}, \bibinfo{author}{Eto, M.},
  \bibinfo{author}{Inoue, D.}, \bibinfo{author}{Nakao, K.},
  \bibinfo{year}{2011}.
\newblock \bibinfo{title}{Statistical analysis of honeypot data and building of
  kyoto 2006+ dataset for nids evaluation}, in: \bibinfo{booktitle}{Proceedings
  of the first workshop on building analysis datasets and gathering experience
  returns for security}, pp. \bibinfo{pages}{29--36}.
\bibitem[{Stolfo et~al.(1999)}]{stolfo1999kdd}
\bibinfo{author}{Stolfo, S.}, et~al., \bibinfo{year}{1999}.
\newblock \bibinfo{title}{Kdd-99 dataset}.
\newblock \bibinfo{journal}{Available on http://www. kdd. ics. uci.
  edu/databases/kddcup99/kddcup99. html kddcup99. html} .
\bibitem[{Sundermeyer et~al.(2012)Sundermeyer, Schl{\"u}ter and
  Ney}]{sundermeyer2012lstm}
\bibinfo{author}{Sundermeyer, M.}, \bibinfo{author}{Schl{\"u}ter, R.},
  \bibinfo{author}{Ney, H.}, \bibinfo{year}{2012}.
\newblock \bibinfo{title}{Lstm neural networks for language modeling}, in:
  \bibinfo{booktitle}{Thirteenth annual conference of the international speech
  communication association}.
\bibitem[{Suykens and Vandewalle(1999)}]{suykens1999least}
\bibinfo{author}{Suykens, J.A.}, \bibinfo{author}{Vandewalle, J.},
  \bibinfo{year}{1999}.
\newblock \bibinfo{title}{Least squares support vector machine classifiers}.
\newblock \bibinfo{journal}{Neural processing letters} \bibinfo{volume}{9},
  \bibinfo{pages}{293--300}.
\bibitem[{Tavallaee et~al.(2009)Tavallaee, Bagheri, Lu and
  Ghorbani}]{tavallaee2009detailed}
\bibinfo{author}{Tavallaee, M.}, \bibinfo{author}{Bagheri, E.},
  \bibinfo{author}{Lu, W.}, \bibinfo{author}{Ghorbani, A.A.},
  \bibinfo{year}{2009}.
\newblock \bibinfo{title}{A detailed analysis of the kdd cup 99 data set}, in:
  \bibinfo{booktitle}{2009 IEEE Symposium on Computational Intelligence for
  Security and Defense Applications}, \bibinfo{organization}{IEEE}. pp.
  \bibinfo{pages}{1--6}.
\bibitem[{Wang et~al.(2017)Wang, Yan, Chen, Yang, Zhao and
  Conti}]{wang2017detecting}
\bibinfo{author}{Wang, S.}, \bibinfo{author}{Yan, Q.}, \bibinfo{author}{Chen,
  Z.}, \bibinfo{author}{Yang, B.}, \bibinfo{author}{Zhao, C.},
  \bibinfo{author}{Conti, M.}, \bibinfo{year}{2017}.
\newblock \bibinfo{title}{Detecting android malware leveraging text semantics
  of network flows}.
\newblock \bibinfo{journal}{IEEE Transactions on Information Forensics and
  Security} \bibinfo{volume}{13}, \bibinfo{pages}{1096--1109}.
\bibitem[{Wang et~al.(2020)Wang, Shang, He, Li and Liu}]{wang2020botmark}
\bibinfo{author}{Wang, W.}, \bibinfo{author}{Shang, Y.}, \bibinfo{author}{He,
  Y.}, \bibinfo{author}{Li, Y.}, \bibinfo{author}{Liu, J.},
  \bibinfo{year}{2020}.
\newblock \bibinfo{title}{Botmark: Automated botnet detection with hybrid
  analysis of flow-based and graph-based traffic behaviors}.
\newblock \bibinfo{journal}{Information Sciences} \bibinfo{volume}{511},
  \bibinfo{pages}{284--296}.
\bibitem[{White(1963)}]{white1963principles}
\bibinfo{author}{White, B.}, \bibinfo{year}{1963}.
\newblock \bibinfo{title}{Principles of neurodynamics: Perceptrons and the
  theory of brain mechanisms}.
\bibitem[{Yu et~al.(2017)Yu, Zhang, Wang and Yu}]{yu2017seqgan}
\bibinfo{author}{Yu, L.}, \bibinfo{author}{Zhang, W.}, \bibinfo{author}{Wang,
  J.}, \bibinfo{author}{Yu, Y.}, \bibinfo{year}{2017}.
\newblock \bibinfo{title}{Seqgan: Sequence generative adversarial nets with
  policy gradient}, in: \bibinfo{booktitle}{Proceedings of the AAAI conference
  on artificial intelligence}.
\bibitem[{Zhou et~al.(2020)Zhou, Cheng, Jiang and Dai}]{zhou2020building}
\bibinfo{author}{Zhou, Y.}, \bibinfo{author}{Cheng, G.},
  \bibinfo{author}{Jiang, S.}, \bibinfo{author}{Dai, M.}, \bibinfo{year}{2020}.
\newblock \bibinfo{title}{Building an efficient intrusion detection system
  based on feature selection and ensemble classifier}.
\newblock \bibinfo{journal}{Computer Networks} , \bibinfo{pages}{107247}.
\bibitem[{Zingo and Novocin(2020)}]{zingo2020can}
\bibinfo{author}{Zingo, P.}, \bibinfo{author}{Novocin, A.},
  \bibinfo{year}{2020}.
\newblock \bibinfo{title}{Can gan-generated network traffic be used to train
  traffic anomaly classifiers?}, in: \bibinfo{booktitle}{2020 11th IEEE Annual
  Information Technology, Electronics and Mobile Communication Conference
  (IEMCON)}, \bibinfo{organization}{IEEE}. pp. \bibinfo{pages}{0540--0545}.

\end{thebibliography}



\end{document}